\documentclass[journal]{IEEEtran}

\usepackage{multirow}
\usepackage[table,xcdraw]{xcolor}
\usepackage{graphicx}
\usepackage{amsmath}
\usepackage{amsfonts}
\usepackage{algorithm}
\usepackage{amsthm,amssymb}
\usepackage{mathrsfs}
\usepackage{algorithmic}
\usepackage{makecell}
\usepackage{easyReview}
\usepackage{booktabs}
\setcellgapes{3pt}
\usepackage{verbatim}
\setcellgapes{3pt}

\usepackage[justification=centering]{caption}
\usepackage{color}
\usepackage[subfigure]{tocloft}
\usepackage{subfigure}
\usepackage{tabularx}
\usepackage{amsmath, bm}
\usepackage{hyperref}
\usepackage{graphicx}

\hypersetup{hidelinks,
	colorlinks=true,
	allcolors=blue,
	pdfstartview=Fit,
	breaklinks=true}

\usepackage{cite}

\ifCLASSINFOpdf

\else

\fi

\begin{document}

\title{FlashSAM: Lightweight Vision Model for Multi-UAV Token Communication in Low-Latitude Wireless Networks}

\author{Feibo Jiang, \textit{Senior Member, IEEE}, Siwei Tu, Li Dong, Kezhi Wang, \textit{Senior Member, IEEE}, Kun Yang, \textit{Fellow, IEEE}, Ruiqi Liu, \textit{Senior Member, IEEE}, Cunhua Pan, \textit{Senior Member, IEEE}, Jiangzhou Wang, \textit{Fellow, IEEE} 
	\thanks{
	This work was supported in part by the National Natural Science Foundation of China under Grants 62572184, 41604117, and 62531008; in part by the Natural Science Foundation of Hunan Province under Grants 2024JJ5270 and 2025JJ50365; in part by the Changsha Natural Science Foundation under Grants kq2402098 and kq2402162; in part by the Jiangsu Major Project on Fundamental Research under Grant BK20243059; in part by the Gusu Innovation Project under Grant ZXL2024360; and in part by the Suzhou High-Tech District under Grant RC2025001. (Corresponding author: Li Dong.)
		
		Feibo Jiang (jiangfb@hunnu.edu.cn) is with the Hunan Provincial Key Laboratory of Intelligent Computing and Language Information Processing and the Yuelushan Digital Intelligence Laboratory (Artificial Intelligence and International Communication), Hunan Normal University, Changsha 410081, China.
		
		Siwei Tu (tusiwei@hunnu.edu.cn) is with the School of Information Science and Engineering, Hunan Normal University, Changsha 410081, China.
		
		Li Dong (Dlj2017@hunnu.edu.cn) is with the School of Computer Science, Hunan University of Technology and Business, Changsha 410205, China, and also with the Xiangjiang Laboratory, Changsha 410205, China.
		
		Kezhi Wang (Kezhi.Wang@brunel.ac.uk) is with the Department of Computer Science, Brunel University London, UB8 3PH London, UK.
		
		Kun Yang (e-mail: kunyang@nju.edu.cn) is with the State Key Laboratory of Novel Software Technology, Nanjing University, Nanjing 210008, China, also with Institute of Intelligent Networks and Communications (NINE) and School of Intelligent Software and Engineering, Nanjing University (Suzhou Campus), Suzhou 215163, China.
		
		Ruiqi Liu (richie.leo@zte.com.cn) is with the Wireless and Computing Research Institute, ZTE Corporation, Beijing 100029, China.
		
		Cunhua Pan (cpan@seu.edu.cn) is with the National Mobile Communications Research Laboratory, Southeast University, Nanjing 210096, China.
		
		Jiangzhou Wang (j.z.wang@seu.edu.cn) is with the National Mobile Communications Research Laboratory, Southeast University, Nanjing, China, and also with the Purple Mountain Laboratories, Nanjing 210096, China.
		
	}

}

\markboth{Submitted for Review}
{Shell \MakeLowercase{\textit{et al.}}: Bare Demo of IEEEtran.cls for IEEE Journals}

\maketitle

\begin{abstract}
Token Communication (TokenCom) is a promising paradigm for low-altitude wireless networks, as it focuses on transmitting task-relevant core information, particularly in environments with uncertainty, noise, and stringent bandwidth constraints. However, existing TokenCom systems still face several challenges, including inefficient knowledge base construction, ineffective token encoding, and limited support for multi-user token sharing. To address these issues, we propose a Lightweight Vision Model-based Multi-Unmanned Aerial Vehicle (UAV) Token Communication (LVM-MTC) system. First, we develop a lightweight Segment Anything Model (SAM), termed FlashSAM, which incorporates a set of lightweight convolutional modules to significantly reduce the number of model parameters. Building on FlashSAM, we construct a Lightweight Knowledge Base (LKB) to enable efficient object-level perception. Next, we design an Efficient Token Codec (ETC) based on the Masked Autoencoder (MAE) architecture. ETC improves compression efficiency at both the pixel and token levels, and provides lightweight token decoding tailored for resource-constrained UAVs. Furthermore, we propose a Multi-UAV Token Sharing (MTS) scheme for multi-UAV TokenCom. By measuring token similarity across UAVs, MTS consolidates similar tokens and transmits them through broadcast transmission, thereby further improving transmission efficiency. Finally, simulation results validate the feasibility and effectiveness of the proposed LVM-MTC system.
\end{abstract}


\section{Introduction}
Low-altitude wireless networks refer to communication networks deployed in low-altitude airspace to support wide-area connectivity for applications such as environmental monitoring, emergency response, aerial sensing, and ubiquitous coverage. With the increasing demand for reliable communications in large-scale and infrastructure-limited environments, these networks play an important role in enabling continuous information acquisition, situational awareness, and coordinated operations across wide geographic areas \cite{wang2015multi}.

Unmanned Aerial Vehicles (UAVs), with their capabilities for rapid deployment, on-demand coverage, and aerial relaying, have become a key enabling technology for low-altitude wireless networks, especially in areas where terrestrial infrastructure is sparse or disrupted. However, stringent constraints on onboard energy, payload, spectrum resources, and computational capacity limit both the available bandwidth and the achievable fidelity of visual data delivery. In downlink missions, base stations (BSs) may need to transmit task-relevant visual content, such as semantic maps, hazard overlays, or shared scene representations for multi-UAV coordination. Nevertheless, directly streaming high-volume images or videos over resource-constrained links is often inefficient and may degrade both timeliness and reliability.

Semantic Communication (SemCom) is a task-oriented paradigm for 6G that transmits task-relevant semantic representations instead of raw bit streams, enabling a shift from signal transmission to meaning transmission while reducing bandwidth consumption and improving energy efficiency and robustness \cite{11224690, 10910032}. Token communication (TokenCom) further represents semantics as compact, discrete tokens and treats them as basic communication units, offering a standardized semantic interface with predictable and rate-controllable transmission. In image communications, Transformer-based receivers can infer and recover missing or uncertain tokens from context, enabling efficient semantic exchange at extremely low bitrates while preserving task-relevant fidelity \cite{11175596}.
However, under limited communication resources and growing user demands, existing image TokenCom systems still face the following challenges:

\subsubsection{Inefficient Knowledge Base}
Traditional SemCom systems often rely on large-scale rule sets or Knowledge Graphs (KGs) as external knowledge bases \cite{shi2021semantic}, whose collection, curation, and updates are costly and complex. From a TokenCom perspective, a lightweight, task-aligned, and tokenizable knowledge interface is needed for efficient querying and invocation during transmission and decoding. Although Large Vision Models (LVMs) contain rich knowledge, their parameter scale and inference latency hinder real-time deployment \cite{wang2024uses,11370829}. Hence, the low-cost and rapid construction of a token-ready knowledge base remains a bottleneck.

\subsubsection{Ineffective Token Encoding}
Images exhibit substantial pixel redundancy and low information density. Convolutional Neural Network (CNN)-based encoders typically encode the whole image, removing redundancy mainly at the semantic level while leaving pixel-level redundancy insufficiently addressed \cite{xie2023communication}. In TokenCom, redundant pixels translate into longer token sequences, weakening rate controllability; over-encoding background regions also reduces the token budget for primary objects and harms task-relevant fidelity. A desirable design should jointly prune pixel redundancy and preserve object-centric semantics to produce compact tokens.

\subsubsection{Lack of Multi-User Token Sharing}
As user populations and application scales grow, point-to-point schemes become inefficient in large-scale multi-user settings. Existing multi-user SemCom studies rarely exploit token-level sharing in the semantic space \cite{xie2022task}. In TokenCom, users’ token streams often overlap in environmental priors, task constraints, and scene elements. By distinguishing shared versus user-specific tokens, the transmitter can merge common parts and send only incremental differences, reducing bandwidth redundancy and improving scalability.

Lightweight vision models have become a key direction for enabling the deployment of LVMs on edge devices \cite{11370176}. To reduce the resource footprint and inference latency of LVMs, researchers have widely adopted techniques such as knowledge distillation, quantization, and pruning to compress LVMs into lightweight models\cite{gou2021knowledge,yao2022zeroquant}. Building on these advances, we propose a Lightweight Vision Model-based Multi-UAV Token Communication System (LVM-MTC). The key innovations of LVM-MTC are as follows:
\setcounter{subsubsection}{0}
\subsubsection{Lightweight Knowledge Base}
To address the high resource consumption and slow inference incurred when LVMs are used to support knowledge interfaces for TokenCom, we design and train a lightweight variant of SAM, termed FlashSAM. By replacing the Vision Transformer (ViT) backbone in SAM with a more compact convolutional structure, FlashSAM achieves an approximately $10\times$ reduction in parameters and an approximately $20\times$ speedup in inference. Building on FlashSAM, we construct a Lightweight Knowledge Base (LKB) that rapidly localizes key semantic objects and their regions, providing a tokenizable prior for subsequent token generation and selection. As a result, the proposed TokenCom system can extract critical tokens from images more quickly and accurately.

\subsubsection{Efficient Token Codec}

To enable more efficient token compression and transmission, we develop an Efficient Token Codec (ETC) based on the Masked AutoEncoder (MAE) architecture. The ETC encoder performs token encoding in two stages. In the low-level (pixel) stage, it leverages the object regions and location cues provided by the LKB to conduct adaptive masking, thereby removing redundant pixels unrelated to the objects. In the high-level (semantic) stage, a ViT-based image encoder generates high-quality tokens and further eliminates redundant semantics. At the receiver, the decoder requires only a small number of tokens to reconstruct object semantics and then recovers fine-grained details by exploiting pixel-level statistical correlations to restore the image. Moreover, ETC adopts an asymmetric design: the computationally intensive encoder is deployed at the BS, whereas the lightweight decoder is deployed on UAVs.
	
\subsubsection{Multi-UAV Token Sharing Transmission}
From the perspective of TokenCom, we exploit the representation power of LVMs to construct a multi-user semantic space, where images from different UAVs are mapped into comparable token representations. Semantically identical or similar tokens are merged into shared tokens. During transmission, the Multi-UAV Token Sharing (MTS) scheme separately conveys shared tokens and UAV-specific private (i.e., differential) tokens to reduce the overall token volume. At the receiver, each UAV recombines the shared and private tokens to reconstruct its corresponding image, thereby enabling a more efficient multi-UAV TokenCom system.

The remainder of this paper is structured as follows: Section \uppercase\expandafter{\romannumeral2}  reviews the related work; Section \uppercase\expandafter{\romannumeral3}  presents the system model; Section \uppercase\expandafter{\romannumeral4}  provides a detailed description of the proposed LVM-MTC; Section \uppercase\expandafter{\romannumeral5} discusses the experimental setup and results; Section \uppercase\expandafter{\romannumeral6} concludes the paper.

\section{Related Work}
\subsection{UAV-based SemCom Systems}
Xu et al. \cite{10422907} proposed a Federated Learning (FL) powered SemCom framework for UAV swarms. To handle dynamic topologies and resource limitations, they introduced a Hierarchical FL architecture featuring centralized intra-cluster training and decentralized inter-cluster aggregation. 
\textcolor{black}{Liu et al. \cite{10678860} introduced a UAV-assisted Mobile Edge Computing (MEC) system that integrates SemCom under jamming attacks. They proposed a Deep Reinforcement Learning (DRL)-based resource management algorithm that jointly optimizes UAV trajectories, user associations, and channel selections to learn and counter the dynamic jamming.}
Hu et al. \cite{10930657} proposed an intelligent resource allocation method for multi-modal SemCom in UAV collaborative relay networks. The framework allows UAVs to dynamically switch between image and text transmission modalities to maximize a novel Quality of Experience (QoE) metric while minimizing cost.

\subsection{\textcolor{black}{Token Communication}}
Qiao et al. \cite{11175596} proposed TokCom, a large-model-driven paradigm for cross-modal SemCom. By representing multimodal data as unified discrete tokens and leveraging pre-trained Multimodal Large Language Models (MLLMs) for masked/next-token prediction, TokCom reconstructs missing tokens at extremely low bitrates, establishing ``tokens as semantics, semantics as communication."
Liu et al. \cite{11149073} designed text-guided TokCom, the first system to employ textual tokens as side information for wireless image transmission. 
Zhang et al. \cite{zhang2025tokcomuepsemanticimportancematchedunequal} presented TokCom-UEP, revealing that equal-error protection wastes redundancy because 1-D token sequences exhibit inherent semantic hierarchy. 
Jiang et al. \cite{jiang2026tokencom} proposed a vision-language-model-driven TokCom paradigm that represents multimodal data as unified semantic tokens for transmission.

\subsection{Multi-user SemCom Systems}
Xie et al. \cite{9830752} studied task-oriented multi-user SemCom systems and proposed a Transformer-based framework to unify the transmitter architecture for different tasks, including image retrieval, machine translation, and visual question answering. 
Li et al. \cite{li2023non} proposed a Non-Orthogonal Multiple Access (NOMA)-based multi-user SemCom system (NOMASC), which supports semantic transmission for multiple users with different source information modalities. The system uses asymmetric quantizers and neural network models for symbol mapping and intelligent multi-user detection. 
Mu et al. \cite{mu2022semantic} introduced an innovative heterogeneous semantic and bit multi-user communication framework that adopts a semi-NOMA scheme to effectively facilitate heterogeneous semantic and bit multi-user communication. They also proposed an opportunistic semantic and bit communication method to alleviate the early and late rate difference issues in NOMA.

Hence, existing UAV-assisted SemCom studies mainly target system-level optimization but often rely on heavy vision models or continuous feature delivery, making realtime semantic perception and transmission difficult under
bandwidth/compute/latency constraints. TokenCom improves
semantic recovery via generative/predictive token completion,
yet its model scale and inference complexity can hinder
deployment on resource-limited UAVs and in low-latency
closed-loop missions. Multi-user SemCom typically separates
users via resource orthogonalization or NOMA-like schemes,
while under-exploiting semantic-space sharing, leaving crossuser redundancy largely untreated at the source. In contrast,
our complexity-aware, edge-friendly design uses FlashSAMenabled LKB for real-time key-object localization, an asymmetric ETC that offloads heavy encoding to the base station
with lightweight UAV decoding, and MTS to merge shared
semantics, reducing redundancy while preserving task-relevant
semantic fidelity.

\section{System Model}
\begin{figure}[htbp]
	\centering
	\includegraphics[width=8cm]{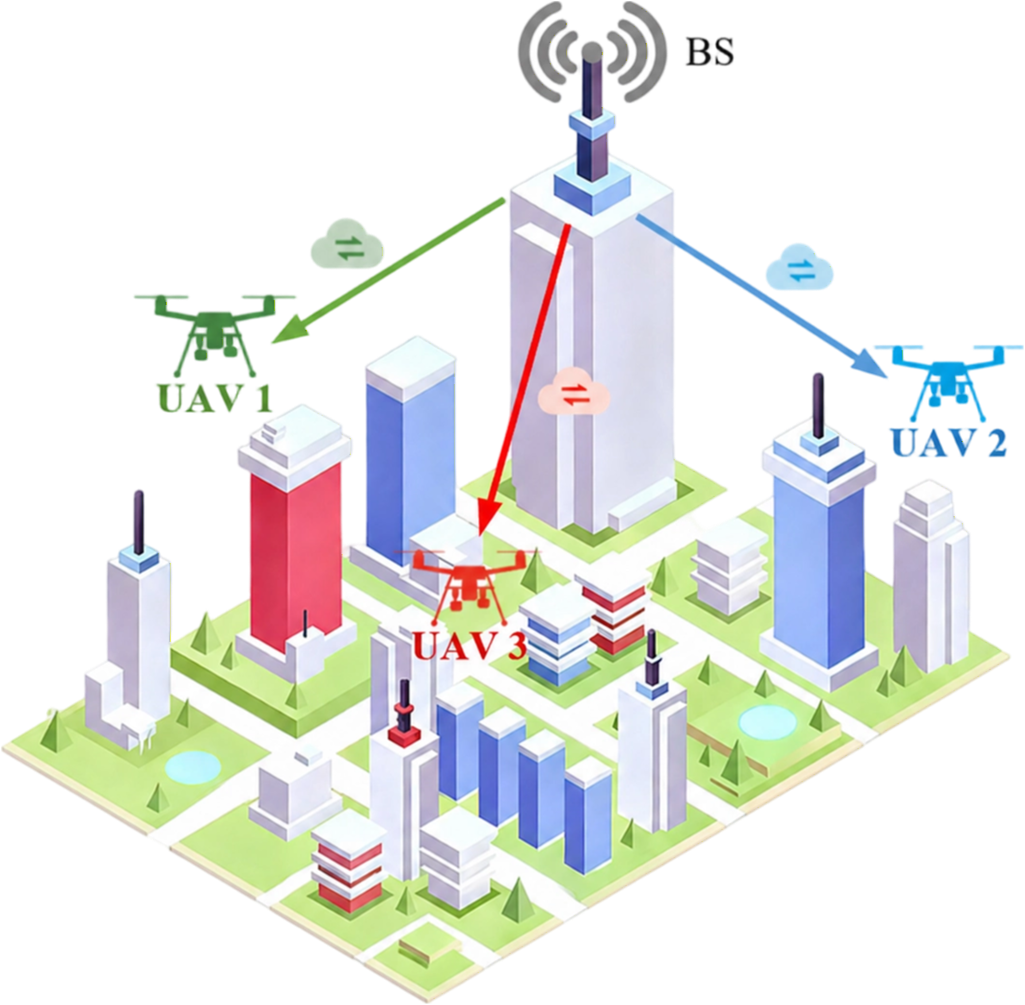}
	\caption{Architecture of the BS-Multi-UAV TokenCom System.}
	\label{fig:uav}
\end{figure}

\begin{figure*}[htbp]
	\centering
	\includegraphics[width=17cm]{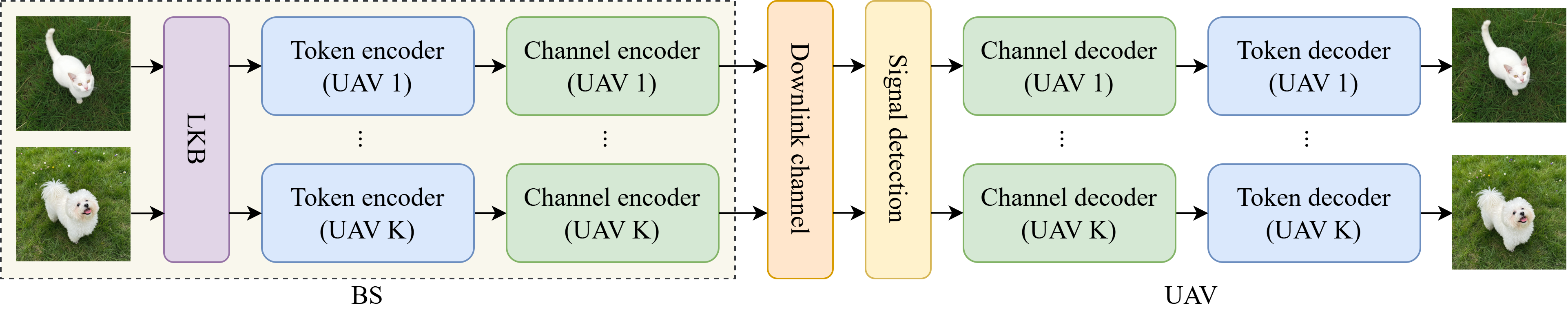}
	\caption{System Model of Downlink Multi-UAV TokenCom.}
	\label{fig:sys}
\end{figure*}

\textcolor{black}{As shown in Fig. \ref{fig:uav} and Fig. \ref{fig:sys}}, we consider a multi-UAV TokenCom system. The system consists of a transmitter (BS) and multiple receivers (UAVs). 
At the BS, $K$ source images are jointly tokenized, and channel encoded, and transmitted over the downlink channel. At the receiver, the received signals are jointly token and channel decoded to reconstruct the source images.

\subsection{BS-Side TokenCom Transmitter}
We denote the source image associated with the $k$-th UAV as ${\textbf P}_k \in \mathbb{R}^{C \times H \times W}$, where $C$, $H$ and $W$ denote the number of channels, height, and width, respectively. Each image contains task-relevant information, which is first transformed into a token representation:
\begin{equation}
	\label{equ:1}
	\textbf Z_k={S\left( {F\left ( {\textbf P}_k;{\theta} \right ) ,{\textbf P}_k ;{\alpha }_k} \right)},
\end{equation}
where ${\textbf Z}_k \in \mathbb{R}^{L_S \times D_S}$ denotes the token representation with token length $L_S$ and embedding dimension $D_S$. Here, ${{S}\left( {;{{{\alpha }}_k}} \right)}$ is the token encoder for the $k$-th UAV with learnable parameters ${{{\alpha }}_k}$, and ${{F}\left( {;{{{\theta }}}} \right)}$ denotes the knowledge base, termed LKB, with fixed parameters $\theta$ shared by all UAVs. Specifically, ${{F}\left( {{\textbf P}_k;{{{\theta }}}} \right)}$ provides the location cues of key task-relevant objects retrieved by the LKB from ${\textbf P}_k$, which are incorporated into tokenization to guide token generation and selection. Due to limited communication resources and complex wireless environments, the token representation of the $k$-th UAV is further compressed and mapped to a complex signal \cite{9830752}:
\begin{equation}
	{\textbf X}_k = C\left( {\textbf Z}_k;{\beta }_k \right),
\end{equation}
where ${\textbf X}_k \in {\mathbb{C} }^{L_S \times D_C}$ is the transmitted complex-valued signal with feature dimension $D_C < D_S$ and ${ C\left( { ;{{{\beta }}_k}} \right)}$ is the channel encoder for the $k$-th UAV with learnable parameters  ${{{\beta }}_k}$. 
After channel encoding, power normalization is performed to satisfy the transmit power constraint \cite{10431795}:
\begin{equation}
	\frac{1}{L_S \times D_C} \mathbb{E}\left[\left\|\textbf{X}_k\right\|_{2}^{2}\right] \leq P_{S},
\end{equation}
where $P_{S}$ denotes the maximum allowable transmit power per transmitter. This step ensures that the power of the transmitted signal
$\textbf{X}_k$ remains within the prescribed limit, enabling efficient utilization of the available transmit power.

\subsection{Downlink Channel}

When the transmitted signal passes through a Multiple-Input Multiple-Output (MIMO) physical channel, the received signal can be expressed as follows \cite{9830752}:
\begin{equation}\label{eq:ch_model}
	\mathbf{Y} = \mathbf{H}\mathbf{X} + \mathbf{N},
\end{equation}
where $\mathbf{X}^T=[\mathbf{X}_1,\mathbf{X}_2,...,\mathbf{X}_K]$ is the transmitted signal for all UAVs, and $\mathbf{H}=[\mathbf{H}_1,\mathbf{H}_2,...,\mathbf{H}_K]$ denotes the channel matrix between the BS and all UAVs. For Rayleigh fading, the entries of $\mathbf{H}$ follow $\mathcal{CN}(0,1)$. $\mathbf{N}$ is the circularly symmetric Gaussian noise with i.i.d. entries $\mathcal{CN}(0,\sigma_n^2)$, and the SNR is defined as $\sum_{k=1}^{K}\left\|\mathbf{H}_k\mathbf{X}_k\right\|^2/\sigma_n^2$.

Subsequently, the transmitted signals are recovered using a Linear Minimum Mean-Squared Error (L-MMSE) detector with the estimated Channel State Information (CSI) \cite{9830752}:
\begin{equation}\label{eq:lmmse_k}
	\widehat{\mathbf{X}}
	=
	\widehat{\mathbf{H}}^{H}
	\left(
	\widehat{\mathbf{H}}\widehat{\mathbf{H}}^{H}
	+\sigma_n^2\mathbf{I}
	\right)^{-1}
	\mathbf{Y},
\end{equation}
where $\widehat{\mathbf{X}}^T=[\widehat{\mathbf{X}}_1; \widehat{\mathbf{X}}_2;...;\widehat{\mathbf{X}}_K]$ is the recovered signal, and $\widehat{\mathbf{H}}=\mathbf{H}+\Delta\mathbf{H}$ denotes the estimated CSI with estimation error $\Delta\mathbf{H}$ whose entries follow $\mathcal{CN}(0,\sigma_e^2)$. Here, $(\cdot)^H$ denotes the Hermitian transpose, $\mathbf{I}$ is the identity matrix, and $\sigma_e^2$ is the CSI error variance.

\subsection{UAV-Side TokenCom Receiver}
The recovered token representation for the $k$-th UAV, $\widehat{\mathbf{Z}}_k \in  {\mathbb{R} }^{L_S \times D_S}$, is obtained by the channel decoder as
\begin{equation}
	\widehat{\mathbf{Z}}_k = C^{-1}\!\left(\widehat{\mathbf{X}}_k;\gamma_k\right),
\end{equation}
where $C^{-1}(\cdot;\gamma_k)$ denotes the channel decoder for the $k$-th UAV with learnable parameters $\gamma_k$. The channel decoder aims to recover the transmitted tokens while mitigating channel distortion and inter-UAV interference. Finally, the token representation is decoded and reconstructed into an image by
\begin{equation}
	\label{equ:7}
	\mathbf{Q}_k = S^{-1}\!\left(\widehat{\mathbf{Z}}_k;\varphi_k\right),
\end{equation}
where $\mathbf{Q}_k$ is the reconstructed image and $S^{-1}(\cdot;\varphi_k)$ denotes the token decoder for the $k$-th UAV with learnable parameters $\varphi_k$.

\subsection{Problem Formulation}
Based on the above system model, we aim to learn a set of TokenCom modules at the BS and UAVs such that the reconstructed images preserve task-relevant information. Specifically, let $\mathcal{L}(\mathbf{P}_k,\mathbf{Q}_k)$ denote the task loss for the $k$-th UAV. With the fixed LKB $F(\cdot;\theta)$, the  problem can be formulated as follows:
\begin{equation}\label{eq:opt_main}
	\begin{aligned}
		\min_{\{\alpha_k,\beta_k,\gamma_k,\varphi_k\}_{k=1}^{K}} \quad 
		& \mathbb{E}_{\{\mathbf{P}_k\}},\mathbb{E}_{\mathbf{H},\mathbf{N}}
		\left[\sum_{k=1}^{K}\mathcal{L}\!\left(\mathbf{P}_k,\mathbf{Q}_k\right)\right] \\	
	\end{aligned}
\end{equation}
where the expectation is taken over the training data distribution, channel fading and additive Gaussian noise.

\section{Proposed LVM-MTC System}
\begin{figure*}[htbp]
	\centering
	\includegraphics[width=16cm]{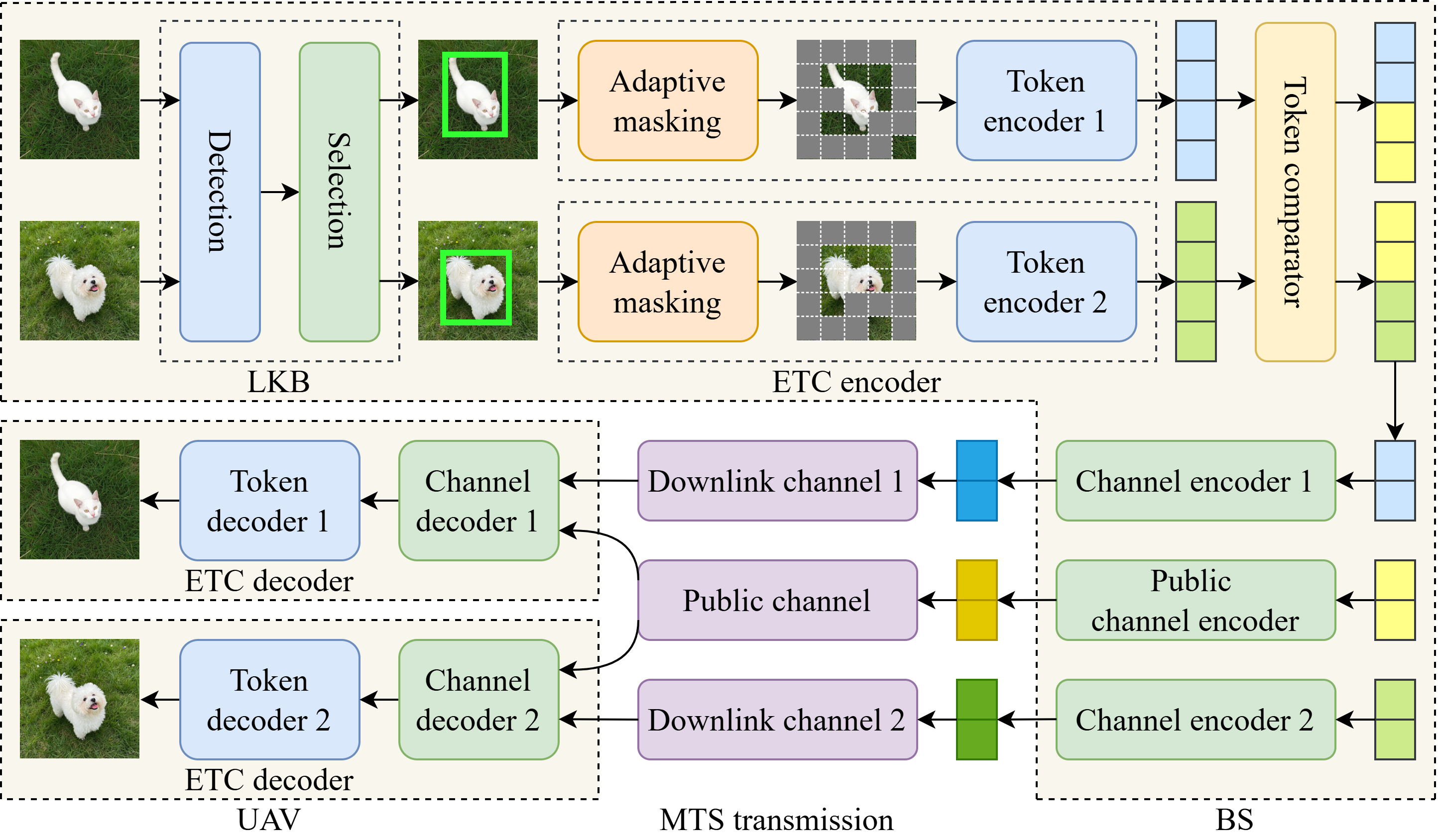}
	\caption{Framework of the Proposed Downlink Multi-UAV TokenCom System.}
	\label{fig:main}
\end{figure*}

In this section, we provide the implementation details of the proposed LVM-MTC system, as illustrated in Fig. \ref{fig:main} (using two UAVs as an example).

\subsection{System Overview}
\subsubsection{LKB}
We first deploy an LKB at the BS, based on the customized FlashSAM model.
The LKB encompasses extensive visual knowledge and can swiftly and accurately locate task-relevant objects for TokenCom in an image \cite{zhao2023fast}. It facilitates the adaptive masking process of the ETC by discarding irrelevant background pixels, thereby improving tokenization efficiency at the pixel level. As illustrated in Fig. \ref{fig:main}, the LKB identifies key objects to be tokenized in different source images by marking them with rectangular bounding boxes.

\subsubsection{ETC Encoder}
We utilize the ETC, consisting of an encoder and decoder, to perform efficient token encoding and decoding. First, the ETC encoder references the location information generated by the LKB to adaptively mask a portion of the source image pixels that are weakly associated with task-relevant objects. Subsequently, the token encoder, composed of multiple Transformer encoder layers, encodes the remaining critical pixels to generate high semantic-density tokens. As shown in Fig. \ref{fig:main}, for a single animal image, background pixels exhibit a weak correlation with the animal itself. Thus, most background pixels are adaptively masked by the ETC, while the ETC encoder performs high-quality token generation on the remaining critical pixels.

\subsubsection{Token Comparator}
At the BS, a token comparator evaluates the similarity of token representations from different UAVs in the semantic space. Each token is divided into two parts: private tokens specific to each UAV and shared tokens common to all UAVs. The shared tokens capture similar features across different UAVs, such as the grassland in the background and the white fur on different animals illustrated in Fig. \ref{fig:main}.

\subsubsection{MTS Transmission}
The shared tokens are encoded using a public channel encoder and transmitted through a public downlink channel to all receiving UAVs, thereby reducing redundant token transmissions. Private tokens, on the other hand, are encoded and transmitted through channel encoders and downlinks corresponding to individual UAVs. At the receiver side, each UAV receives both the shared and private tokens. These are then combined and decoded using the UAV-specific channel decoder.

\subsubsection{Channel Encoder and Decoder}
The encoded tokens are processed through a channel encoder to ensure effective transmission over the physical channel. The channel codec consists of an autoencoder, where the feature dimensions of the hidden layers progressively decrease to achieve data compression. A perceptron neural network is used to simulate the physical channel, which essentially represents a mapping between transmitted and received signals. This mapping is determined by the adopted channel model (e.g., additive Gaussian noise and fading), rather than trainable weights and biases, and it is introduced to enable end-to-end learning of robust tokens under channel distortions. At the receiver side, a channel decoder is employed to decode the transmitted tokens. To maintain consistency, the channel decoder adopts an autoencoder structure that is the inverse of the channel encoder.

\subsubsection{ETC Decoder}
At the receiver side, the ETC decoder, composed of multiple Transformer decoder layers, functions as a token decoder to reconstruct the original image from the limited received tokens. It is worth noting that the structures of the token encoder and decoder are asymmetrical. The token encoder, which produces compact tokens, has a larger number of parameters and is deployed at the BS, whereas the token decoder, responsible for reconstructing the image, has fewer parameters and is deployed on the UAVs.

\subsection{Lightweight Knowledge Base}

\begin{figure}[htbp]
	\centering
	\includegraphics[width=8cm]{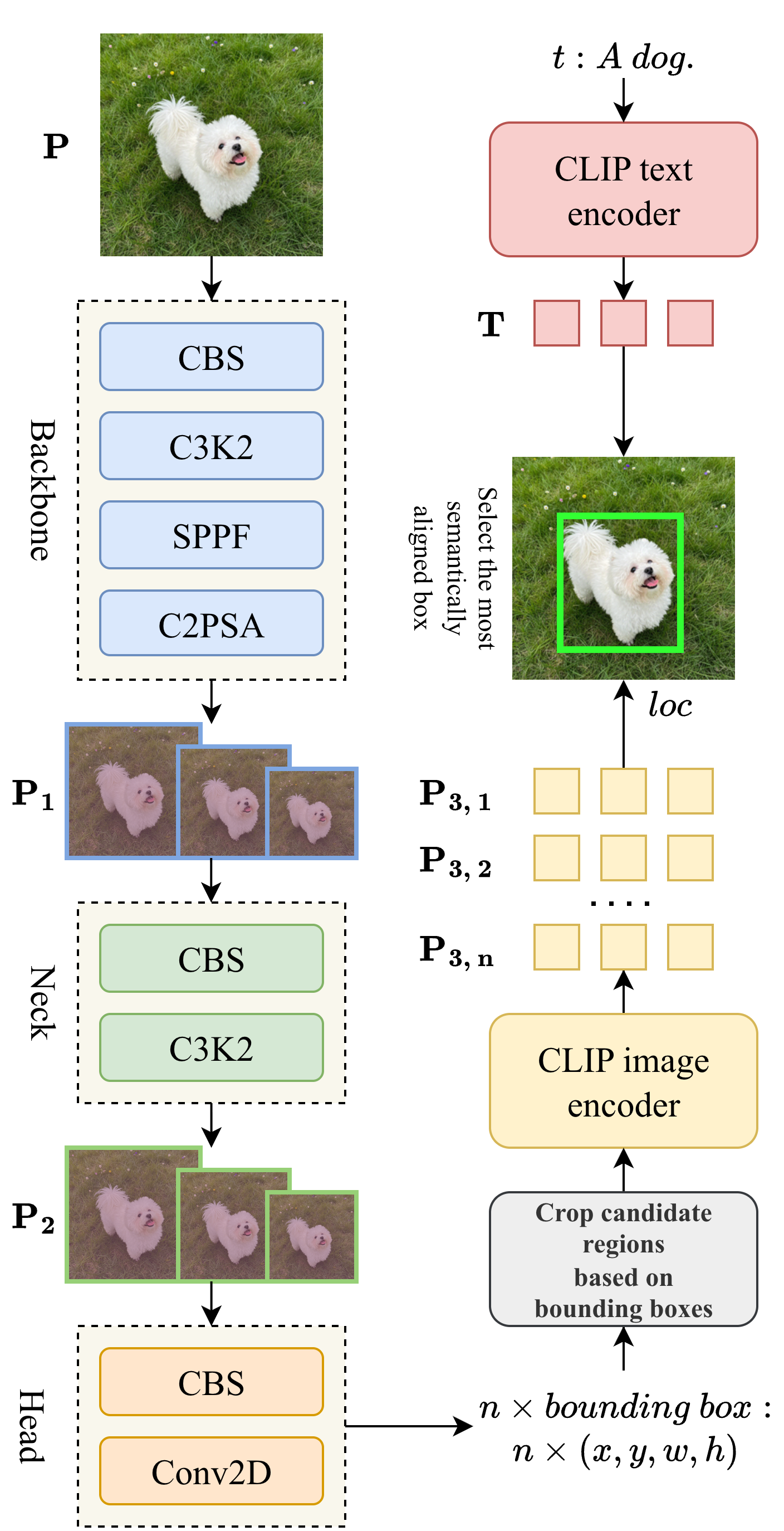}
	\caption{\textcolor{black}{Workflow of FlashSAM (LKB).}}
	\label{fig:msam}
\end{figure}

\textcolor{black}{As illustrated in Fig.~\ref{fig:msam}, the LKB serves as a perception-driven spatial prior generator for the ETC encoder. It leverages FlashSAM to extract object-level bounding boxes for subsequent adaptive masking. Compared with the original SAM that relies on a heavy Vision Transformer encoder, FlashSAM adopts a convolution-dominant architecture, which significantly reduces computational overhead while maintaining accurate object localization. This design makes FlashSAM suitable for real-time UAV-assisted TokenCom. FlashSAM’s architecture is as follows:}

\subsubsection{\textcolor{black}{Backbone}}

\textcolor{black}{FlashSAM first extracts hierarchical visual representations from the image $\mathbf{P}\in\mathbb{R}^{C\times H\times W}$ through a lightweight convolution-dominant backbone. The multi-scale feature maps are obtained as follows \cite{khanam2024yolov11overviewkeyarchitectural}:}
\begin{equation}
	\label{equ:11}
	\color{black}
	\mathbf{P}_{1}=\mathrm{C2PSA}(\mathrm{SPPF}(\mathrm{C3K2}(\mathrm{CBS}(\mathbf{P})))),
\end{equation}
\textcolor{black}{where $\mathbf{P}_{1}$ denotes the multi-scale backbone representation (including small, medium, and large-scale feature responses). $\mathrm{CBS}(\cdot)$, $\mathrm{C3K2}(\cdot)$, $\mathrm{SPPF}(\cdot)$ and $\mathrm{C2PSA}(\cdot)$ constitute the backbone as efficient convolution-based modules: $\mathrm{CBS}(\cdot)$ serves as a basic Conv--BN--SiLU block for local feature extraction; $\mathrm{C3K2}(\cdot)$ is a lightweight residual unit inspired by the Cross Stage Partial (CSP) architecture, designed to enhance gradient flow while reducing computational complexity; $\mathrm{SPPF}(\cdot)$ applies multi-kernel pooling to enlarge receptive fields and capture multi-scale context; and $\mathrm{C2PSA}(\cdot)$ employs parallel split attention to strengthen informative channels and suppress redundancy.}

\subsubsection{\textcolor{black}{Neck}}

\textcolor{black}{To enhance scale robustness and contextual consistency, a feature aggregation neck performs both top-down and bottom-up fusion on $\mathbf{P}_{1}$, which can be expressed as follows:}
\begin{equation}
	\label{equ:12}
	\color{black}
	\mathbf{P}_{2}=\mathrm{C3K2}(\mathrm{CBS}(\mathbf{P}_{1})),
\end{equation}
\textcolor{black}{where $\mathbf{P}_{2}$ denotes the resulting feature representation, which is enriched with multi-scale context and serves as the input to the prediction head.}

\subsubsection{\textcolor{black}{Head}}

\textcolor{black}{The prediction head maps the fused features to object-level bounding boxes and corresponding confidence scores:}
\begin{equation}
	\label{equ:14}
	\color{black}
	\{\mathbf{b}_i, s_i\}_{i=1}^{N}
	=\mathrm{Conv2D}(\mathrm{CBS}(\mathbf{P}_{2})),
\end{equation}
\textcolor{black}{where $\mathbf{b}_i=(x_i,y_i,w_i,h_i)$ denotes the center coordinates, width, and height of the $i$-th candidate bounding box, $s_i$ is the corresponding objectness confidence score, and $N$ is the total number of predicted candidates. Here, $\mathrm{Conv2D}(\cdot)$ denotes the final prediction convolution that maps refined features to detection outputs.}

\subsubsection{\textcolor{black}{Goal-oriented filtering}}

\textcolor{black}{To align perception outputs with the downstream semantic task, we perform text-guided filtering using the shared task description $t$. First, each candidate region $\mathbf{b}_i$ is cropped from $\mathbf{P}$ and encoded to obtain its visual embedding $\mathbf{e}_i$ using the CLIP image encoder. The task text $t$ is encoded using the CLIP text encoder to produce the semantic embedding $\mathbf{e}_t=\mathrm{CLIP}(t)$ \cite{radford2021learning}.}
\textcolor{black}{The cosine similarity between the $i$-th candidate and the task text is defined as follows:}
\begin{equation}
	\label{equ:15}
	\color{black}
	\mathrm{Sim}_i
	=
	\frac{\mathbf{e}_i^{\top}\mathbf{e}_t}
	{\|\mathbf{e}_i\|_2\,\|\mathbf{e}_t\|_2},
	\quad i=1,2,\ldots,N.
\end{equation}

\textcolor{black}{To obtain a compact and precise spatial prior, we select the most semantically aligned candidate and use its bounding box as $loc$:}
\begin{equation}
	\label{equ:16}
	\color{black}
i^\star = \arg\max_i \mathrm{Sim}_i, \quad \mathrm{loc} = b_{i^\star},
\end{equation}
\textcolor{black}{where $\mathbf{b}_{i^{\star}}$ is the corresponding bounding box selected as the final location prior.}

\textcolor{black}{Finally, we adopt FlashSAM as the LKB.  Given an input image $\mathbf{P}$, the LKB produces a task-aligned spatial prior as follows:}
\begin{equation}
\textcolor{black}{loc = \mathcal{F}_{\mathrm{LKB}}(\mathbf{P}, t).}
\end{equation}

\textcolor{black}{The resulting $loc$ is fed into the ETC encoder to guide adaptive token-aware masking, thereby suppressing irrelevant regions and preserving task-critical content for TokenCom. Through this mechanism, the LKB transforms generic object-detection outputs into structured, tokenizable priors.}
\subsection{Efficient Token Codec}
The MAE is a vision model that is pretrained via self-supervised learning on large-scale image data. During the encoding phase, the input image is divided into several patches, many of which are randomly masked. The encoder only processes a small number of patches and attempts to reconstruct the image during the decoding phase \cite{he2022masked}. As a result, MAE achieves a high compression ratio when extracting compact token representations from the image.

\begin{figure}[htbp]
	\centering
	\includegraphics[width=7cm]{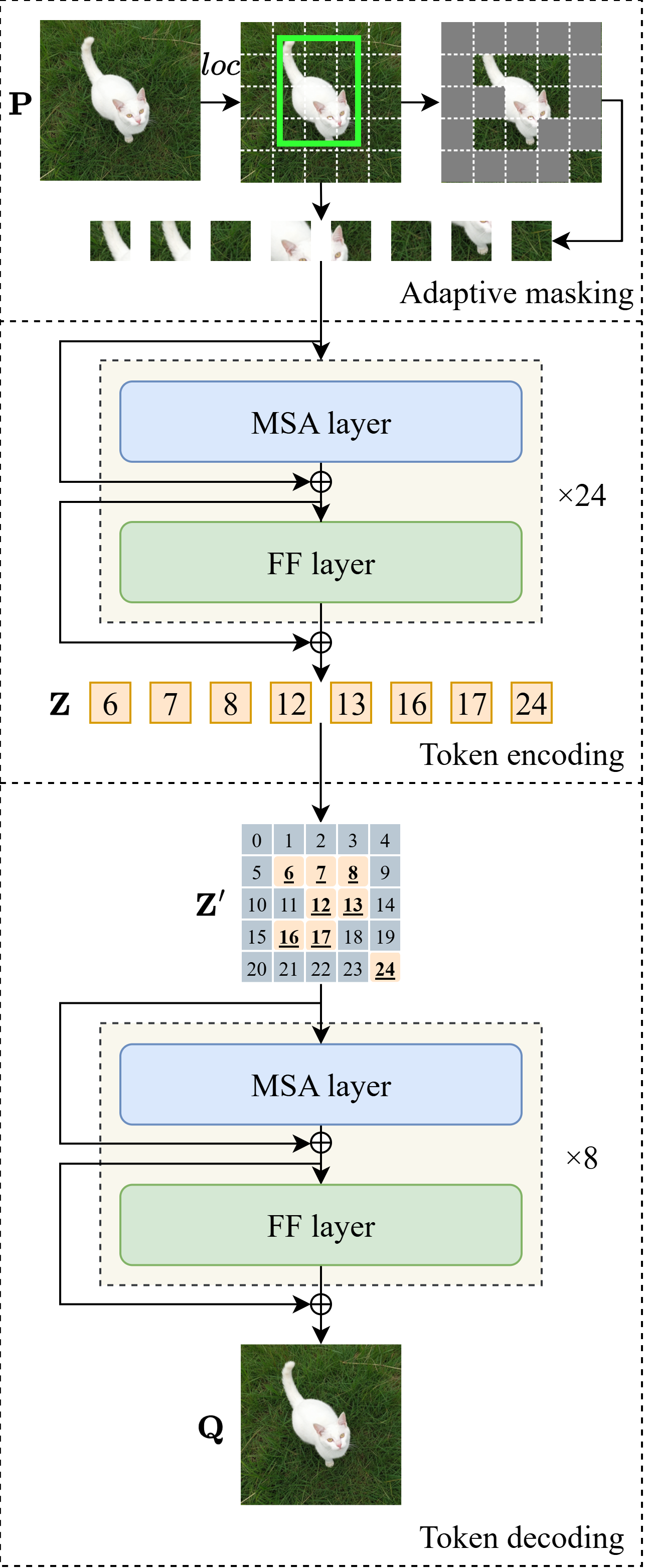}
	\caption{\textcolor{black}{Workflow of ETC. The image is split into patches in a predefined order. Patches whose center points fall inside bounding boxes are treated as foreground and assigned a lower masking probability, while the others are assigned a higher masking probability. Only unmasked patches are encoded. During decoding, masked positions share one mask token, which is combined with encoded features to reconstruct the image.}}
	\label{fig:mae}
\end{figure}

In the proposed LVM-MTC, we develop an efficient token codec, termed ETC, based on MAE. ETC uses the object-location priors provided by the LKB to replace the random masking strategy in MAE with targeted masking, thereby preferentially retaining object-related patches and generating more informative tokens under the same token budget. In this way, more pixels involved in token encoding are associated with task-relevant objects, enabling efficient tokenization and reliable TokenCom transmission. The workflow is shown in Fig.~\ref{fig:mae}. Given an image $\mathbf{P}$ and the location prior $loc$ from the LKB, ETC encodes tokens $\mathbf{Z}$ and reconstructs the image $\mathbf{Q}$ through token decoding. The process is detailed as follows.

\subsubsection{Adaptive Masking}
First, we incorporate the location prior \(loc\) provided by the LKB to prioritize masking the background pixels of \(\textbf{P}\). This removes redundant pixels that are weakly related to target objects, thereby improving token encoding efficiency.
Specifically, the LKB outputs the bounding box coordinates of the selected task-relevant objects. To align these continuous coordinates with the discrete patch-based MAE, we adopt a center-point assignment rule: for each patch, we compute its geometric center; if the center lies inside an LKB bounding box, the patch is classified as a foreground patch (i.e., within the region $loc$); otherwise, it is classified as a background patch.
It is important to note that this does not imply completely discarding background pixels during encoding. Instead, background patches are assigned a higher probability of being masked, while foreground patches are less likely to be masked, so that the encoder can still preserve context tokens when needed.
The implementation details are as follows: suppose the image \(\textbf{P}\) is divided into multiple patches, with the indices represented by \(\mathcal{I} = \{1, 2, \ldots\}\). Therefore, our masking strategy can be expressed as Eq. (\ref{equ:17}), and \( P_i \) is the probability that the \( i \)-th patch is masked:

\begin{equation}
	\label{equ:17}
	P_i = 
	\begin{cases} 
		P_r, & i \in R \\ 
		1 - P_r, & i \in R^c 
	\end{cases},
\end{equation}
where \( P_r \) represents the probability that a foreground patch in \( loc \) is masked. \(R\) is the set of image patches located in \( loc \) and \(R^c\) is the set of image patches located outside \(loc\). Since the goal of LVM-MTC is to transmit object-centric tokens with higher accuracy, \( P_r \) should be set to a value less than 0.5. This makes background patches more likely to be masked, thereby improving the quality of the token representation.
Therefore, the masking indicator of each patch follows a Bernoulli distribution\cite{weisstein2002bernoulli}, which determines whether the patch is masked.
A single forward-pass sampling step for each patch determines the final kept-token set $\mathcal{I}_{keep}$.

\subsubsection{Token Encoding}
The token encoder–decoder framework in LVM-MTC consists of multiple Multi-Head Self-Attention (MSA) layers and Feed-Forward (FF) layers \cite{vaswani2017attention}. Specifically, the token encoder, deployed at the BS, stacks 24 layers to learn hierarchical visual features and generate high-quality tokens. In contrast, the token decoder, deployed on UAVs, stacks only 8 layers, making it significantly lighter and reducing the deployment burden on edge devices. The MSA layer captures global dependencies among different positions in the patch sequence, while the FF layer processes each position independently to model fine-grained local patterns \cite{vaswani2017attention}. Let the set of unmasked patches used for token encoding and their corresponding index set be denoted as \( \textbf{P}_{keep} \) and \( \mathcal{I}_{keep} \), respectively. The outputs of the MSA and FF layers can be expressed as follows:
\begin{equation}
	\label{equ:18}
\textbf{M} _{msa, i}=\left\{\begin{array}{cc}\mathrm{MSA} \left({\mathrm{LN} }\left(\textbf{P} _{keep}\right)\right)+\textbf{P} _{keep}, & i=1 \\\mathrm{MSA} \left({\mathrm{LN} }\left(\textbf{M} _{ff, i-1}\right)\right)+\textbf{M} _{ff, i-1}, & 2 \leq i \leq 24	\end{array}, \right.
\end{equation}
\begin{equation}
	\label{equ:19}
	\textbf{M} _{ff, i}=\mathrm{GeLU}\left(\textbf{W} _{b,f} \cdot \mathrm{LN}\left(\textbf{M} _{msa, {i}}\right)+\textbf{b} _{b, f}\right)+\textbf{M} _{msa, i},
\end{equation}
where \( \text{MSA}(\cdot) \) represents the MSA operator, and \( \text{LN}(\cdot) \) denotes layer normalization. \( \textbf{W}_{b,f} \) and \( \textbf{b}_{b,f} \) are the weights and biases of the FF layer, respectively. \( \text{GeLU}(\cdot) \) denotes the activation function. Finally, the output of the token encoder can be expressed as follows:
\begin{equation}
	\label{equ:20}
	\textbf{Z}=\mathrm{LN}\left ( \textbf{M} _{ff, 24} \right ) .
\end{equation}

\subsubsection{Token Decoding}
\textcolor{black}{At the receiver side, the token decoder reconstructs the image \( \textbf{Q} \) from the channel-decoded tokens \( \hat{\textbf{Z}} \). Following the MAE decoding principle, a shared learnable mask token is used for all masked patches: the decoder copies this mask token to the positions indexed by \( \mathcal{I}-\mathcal{I}_{{keep}} \), concatenates it with the received tokens \( \hat{\textbf{Z}} \), and forms a full-length token sequence for decoding:}
\begin{equation}
	\label{equ:21}
	\textcolor{black}{\hat{\textbf{Z} }^{\prime}=\mathrm{Concat}\left(\hat{\textbf{Z} },\mathrm{Copy}(\mathrm{\textbf{Z}_{mask}},\mathcal{I}-\mathcal{I}_{keep})\right),}
\end{equation}
\textcolor{black}{where \(\textbf{Z}_{mask}\) is the shared mask token, \( \mathrm{Copy}(\cdot) \) denotes copying the mask token to all masked positions, and \( {\hat{\textbf{Z}}}^\prime \) is the full token sequence used for token decoding.}
Finally, the token decoder then continues to decode \( {\hat{\textbf{Z}}}^\prime \) to generate \( \textbf{Q} \)
\begin{equation}
	\label{equ:22}
	\textbf{M} _{msa, i}^\prime=\left\{\begin{array}{cc}\mathrm{MSA} \left({\mathrm{LN} }\left({\hat{\textbf{Z}}}^\prime\right)\right)+{\hat{\textbf{Z}}}^\prime, & i=1 \\\mathrm{MSA} \left({\mathrm{LN} }\left(\textbf{M} _{ff, i-1}^\prime \right)\right)+\textbf{M} _{ff, i-1}^\prime, & 2 \leq i \leq 8	\end{array},\right.
\end{equation}
\begin{equation}
	\label{equ:23}
	\textbf{M} _{ff, i}^\prime=\mathrm{GeLU}\left(\textbf{W} _{b,f} \cdot \mathrm{LN}\left(\textbf{M} _{msa, {i}}^\prime\right)+\textbf{b} _{b, f}\right)+\textbf{M} _{msa, i}^\prime,
\end{equation}
\begin{equation}
	\label{equ:24}
	\textbf{Q}=\mathrm{LN}\left ( \textbf{M} _{ff, 8}^\prime \right ) .
\end{equation}
where $\mathbf{M}_{\mathrm{msa},i}^\prime$ and $\mathbf{M}_{\mathrm{ff},i}^\prime$ are the outputs of the $i$-th MSA and FF layers in the decoder, respectively, with $i=1,2,\ldots,8$. The reconstructed image is denoted by $\mathbf{Q}$.

\subsection{Multi-UAV Token Sharing Transmission}
\begin{figure*}[htbp]
	\centering
	\includegraphics[width=17cm]{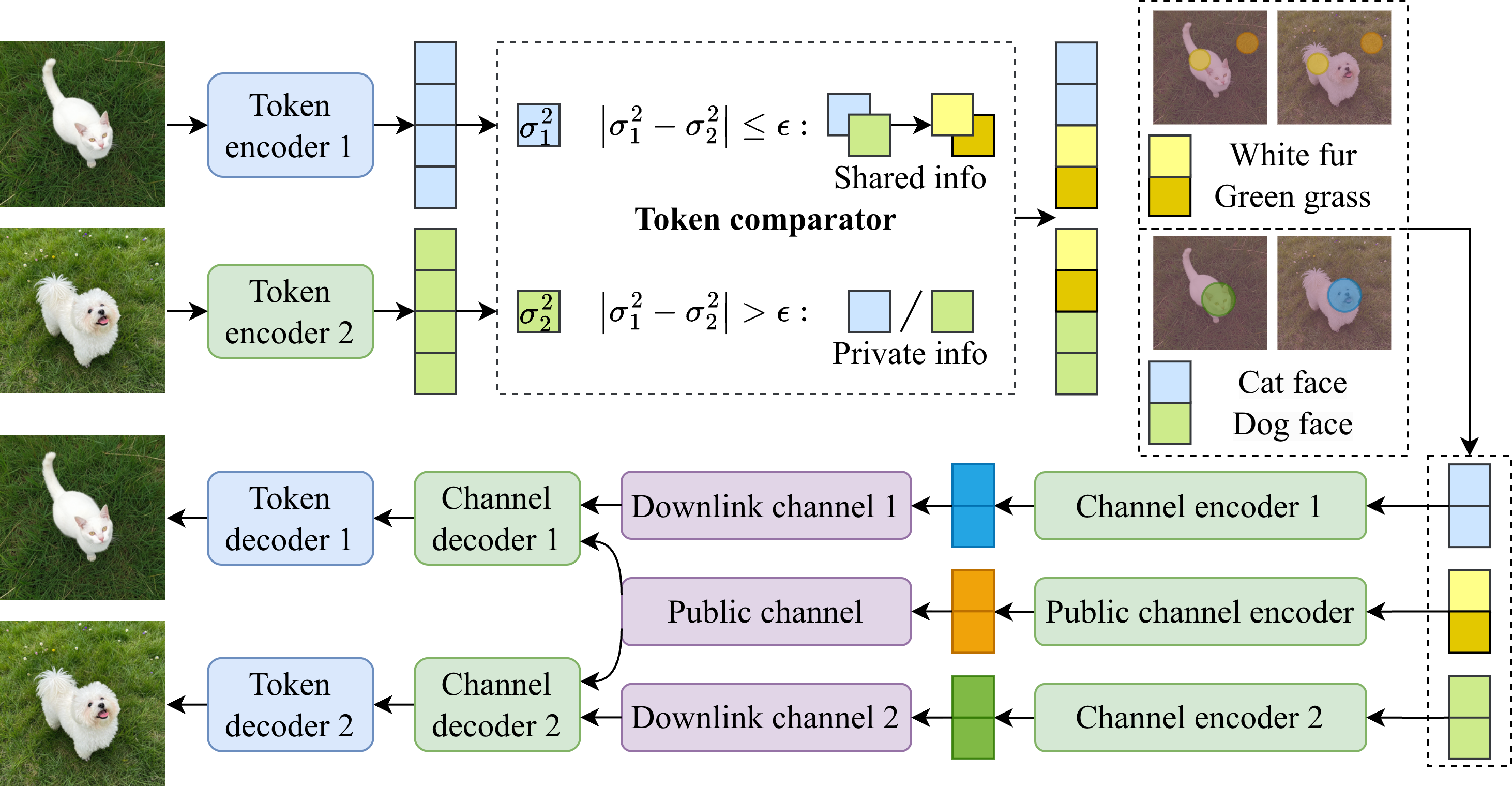}
	\caption{\textcolor{black}{Token Sharing Transmission for Multi-UAV Communication Systems.}}
	\label{fig:mdma}
\end{figure*}

Numerous studies have demonstrated that the semantic information of different images can exhibit significant similarities \cite{zhang2023model}. For example, as shown in Fig. \ref{fig:mdma}, the grass background and the white fur pixel regions in two images share token-level similarity. The proposed image TokenCom system leverages the precise perception and representation capabilities of LVMs to map the source images of different UAVs into a shared token space. A token comparator is then employed in this space to identify, analyze, and extract shared tokens and private tokens from the token representations of multiple UAVs. By modeling UAV-specific characteristics, the system differentiates the private token components across UAVs. During wireless transmission, signals containing shared tokens are broadcast and reused, thereby reducing the overall amount of information transmitted.

Taking the TokenCom system with two UAVs as an example, the left part of Fig. \ref{fig:mdma} illustrates how the token comparator extracts shared and private tokens from the token representations of multiple UAVs, while the right part of Fig. \ref{fig:mdma} provides intuitive examples showing the meanings of shared and private token components. The specific steps are as follows.

\subsubsection{Token Encoding}
As shown in Fig. \ref{fig:mdma}, each UAV independently utilizes a token encoder \(S\left ( ;\alpha _k \right ) \) to extract tokens \(\textbf{Z}_k \in \mathbb{R}^{L_S \times D_S}\) from the source image \(\textbf{P}_k\), as described in Eq. (\ref{equ:1}). Each vector \(\textbf{Z}_{k,i} \ (i = 1, 2, \ldots, L_S)\) in \(\textbf{Z}_k\) can be interpreted as the token embedding corresponding to a specific image patch.

\subsubsection{Token Comparison}

Certain regions across different images often exhibit similar visual patterns, providing a basis for extracting public tokens in multi-UAV TokenCom.
Inspired by MDMA \cite{zhang2023model, 10679082}, we evaluate token commonality based on activation-intensity consistency rather than Euclidean distance or cosine similarity. This method better aligns with the sharing-oriented redundancy reduction objective of TokenCom, as it can preferentially identify tokens that recur stably across users and reduce redundant transmissions at the source. Meanwhile, this statistics-based approach is lightweight and more robust to scale variations and noise, making it suitable for resource-constrained multi-UAV scenarios.

 Specifically, each UAV generates a set of tokens from its image, and we assign each token an intensity statistic $\sigma_k^2$ to characterize its activation strength or information concentration in the high-dimensional token space. We then compare these statistics across UAVs for candidate token pairs and aggregate the discrepancies over all UAV pairs to obtain a token-level discrepancy score. If the score falls below a preset threshold $\epsilon$, the corresponding tokens are deemed sufficiently consistent and are classified as public tokens; otherwise, they are treated as private tokens.

For public tokens, we further construct a compact public representation by averaging the features of the matched public tokens across UAVs, yielding shared tokens that can be broadcast and reused. In contrast, private tokens preserve UAV-specific details and are transmitted separately to each UAV. As illustrated in Fig. \ref{fig:mdma}, public tokens typically capture common patterns across sources (e.g., shared grassy background or similar white-fur regions), whereas private tokens reflect source-specific semantics (e.g., distinct facial structures of the cat and dog). In summary, the public and private tokens obtained by the token comparator are denoted as
\begin{equation}
	\label{equ:30}
	\{\textbf{Z}_{pub},\textbf{Z}_{pri}\}=S_c(\textbf{Z} ),
\end{equation}
where \(\textbf{Z} = \left[ \textbf{Z}_1, \textbf{Z}_2, \ldots, \textbf{Z}_K \right] \in \mathbb{R}^{K \times L_s \times D_s}\) represents the tokens of all UAVs. \(S_c\) is the token comparator. \(\textbf{Z}_{pub}\) represents the public token set shared across the UAVs, whereas \(\textbf{Z}_{pri} = \left[ \textbf{Z}_{pri,1}, \textbf{Z}_{pri,2}, \ldots, \textbf{Z}_{pri,K} \right] \) contains UAV-specific private tokens that capture individualized content. 

\subsubsection{Channel Encoding}
As shown in Fig. \ref{fig:main}, after token encoding and token comparison are completed, the private tokens are encoded by UAV-specific channel encoders, while the public tokens are encoded by an additional public (broadcast) channel encoder for broadcast reuse:
\begin{equation}
	\textbf{X} _{pub}=C_{pub}(\textbf{Z} _{pub};\beta _{pub}),
\end{equation}
\begin{equation}
	\textbf{X} _{pri,k}=C(\textbf{Z} _{pri,k};\beta _k),
\end{equation}
where \(C_{pub}\) represents the public channel encoder, and \(C(;\beta _k)\) denotes the channel encoder for the \(k\)-th UAV. 
Accordingly, $\textbf{X}_{pub}$ is transmitted once and reused by all UAVs, whereas $\textbf{X}_{pri,k}$ delivers only the token differentials required by the $k$-th UAV, thereby reducing redundant token transmissions.

\subsubsection{Channel Decoding}
Subsequently, \(\textbf{X}_{pub}\) is transmitted to all receiving UAVs through a shared channel, while \(\textbf{X}_{pri,k}\) is transmitted to the \(k\)-th UAV through a dedicated channel. Upon reception, each UAV utilizes a channel decoder to recover the combination of public and private tokens.
\begin{equation}
	\label{equ:33}
	\hat{\textbf{Z} }_k=C^{-1}\left(\mathrm{Cat}(\hat{\textbf{Z} }_{pub},\hat{\textbf{Z} }_{pri,k});\gamma_k\right),
\end{equation}
where \(C^{-1}(;\gamma_k)\) is the channel decoder for the \(k\)-th UAV. $\mathrm{Cat}(\cdot)$ denotes concatenation.
The recovered token \(\hat{\textbf{Z}}_k\) is decoded by the token decoder \(S^{-1}(;\varphi_k)\) to reconstruct the image \(\textbf{Q}_k\) as shown in Eq. (\ref{equ:7}).

\subsection{\textcolor{black}{TokenCom Training}}
\begingroup\color{black}
We define the overall training objective of LVM-MTC as minimizing the reconstruction distortion between the input image and its reconstruction, while ensuring reliable token recovery under channel impairments. The training pipeline includes four steps \cite{11175646}.

\subsubsection{FlashSAM Training}We first train FlashSAM to serve as the LKB. FlashSAM adopts a YOLOv11x-seg style convolutional backbone design \cite{khanam2024yolov11}. Specifically, FlashSAM is trained from random initialization on 2\% of the SA-1B dataset for 100 epochs, and the resulting model achieves downstream performance comparable to SAM \cite{kirillov2023segment}. After convergence, FlashSAM is frozen and used only for inference to generate task-aligned spatial priors, without participating in parameter updates during the subsequent TokenCom training.

\subsubsection{Token Codec Training} With the LKB fixed, we train the token encoder-decoder to reconstruct images using the Mean Squared Error (MSE) loss as follows:
\begin{equation}
	\mathcal{L}_{\mathrm{MSE,1}}=\mathbb{E}\!\left[\left\|\mathbf{Q}_k-\mathbf{P}_k\right\|_2^2\right],
\end{equation}
which encourages learning compact token representations that preserve task-relevant content.

\subsubsection{Channel Codec Training} We then train the channel encoder-decoder to robustly compress and recover tokens under the downlink channel, using the MSE loss function:
\begin{equation}
	\mathcal{L}_{\mathrm{MSE,2}}=\mathbb{E}\!\left[\left\|\hat{\mathbf{Z}}_k-\mathbf{Z}_k\right\|_2^2\right].
\end{equation}

\subsubsection{End-to-end Fine-tuning} Finally, we jointly fine-tune the TokenCom system (excluding the frozen LKB) with the loss function as follows:
\begin{equation}
	\mathcal{L}_{\mathrm{MSE,3}}=\mathbb{E}\!\left[\left\|\mathbf{Q}_k-\mathbf{P}_k\right\|_2^2+\left\|\hat{\mathbf{Z}}_k-\mathbf{Z}_k\right\|_2^2\right].
\end{equation}

This design achieves a balanced optimization of reconstruction quality and transmission robustness.

\endgroup

\begin{figure*}[htbp]
	\centering
	\includegraphics[width=17cm]{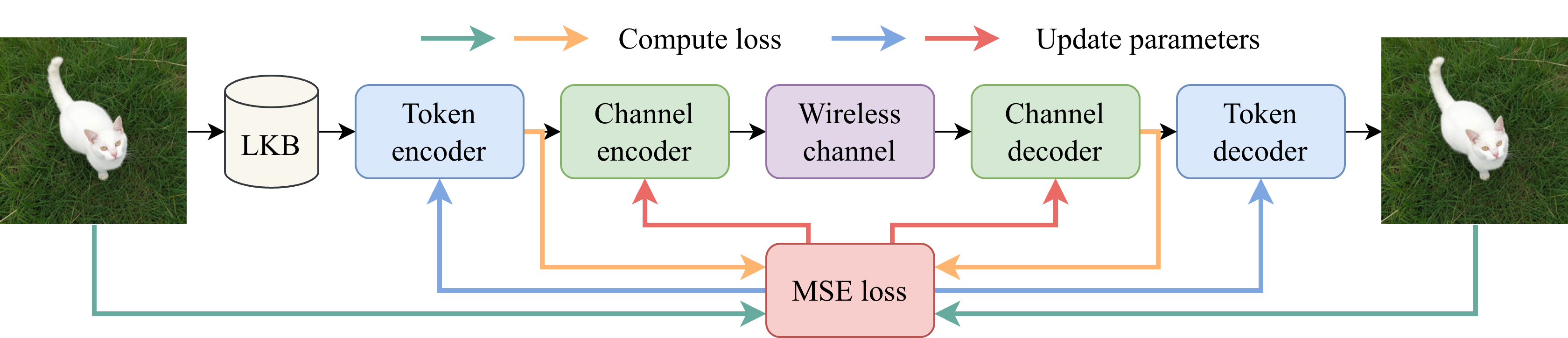}
	\caption{Training of the proposed TokenCom system.}
	\label{fig:train}
\end{figure*}

\section{Simulation Results}

\subsection{Simulation Settings}
The image datasets used in this study include PASCAL VOC2012 and CIFAR-10 \cite{everingham2010pascal}, which contains over 12,000 images spanning 20 categories of everyday objects such as humans, vehicles, and animals. During the experiments, the dataset is evenly partitioned to emulate UAV-specific image sources in the multi-UAV TokenCom setting.

The detailed experimental configuration is as follows: The LKB consists of 56M parameters. The efficient token encoder, based on MAE, includes 24 layers of Vision Transformer with a feature dimension of 1024 \cite{he2022masked}. The channel encoder is composed of a linear layer with an input feature dimension of 1024 and an output feature dimension of 128. To maintain consistency, the channel decoder adopts a structure inverse to the channel encoder. The token decoder, based on the MAE decoder, consists of 8 layers of Vision Transformer. During adaptive masking of ETC, the default masking probability \(P_r\) for object patches is set to 0.25. The threshold \(\epsilon\) for the token comparator is set to 0.1 by default. The channel codec used in the experiments is trained for 10 epochs with a learning rate of \(2 \times 10^{-4}\).    

The training and testing environment includes Python 3.8, PyTorch 2.0.1, and CUDA 11.8. The computational resources are provided by a 12th Gen Intel(R) Core(TM) i7-12700H 2.30 GHz CPU and an A800 GPU with 80 GB of memory.  

\subsection{Performance Evaluation for LKB}
\begin{figure}[htbp]
	\centering
	\includegraphics[width=8.5cm]{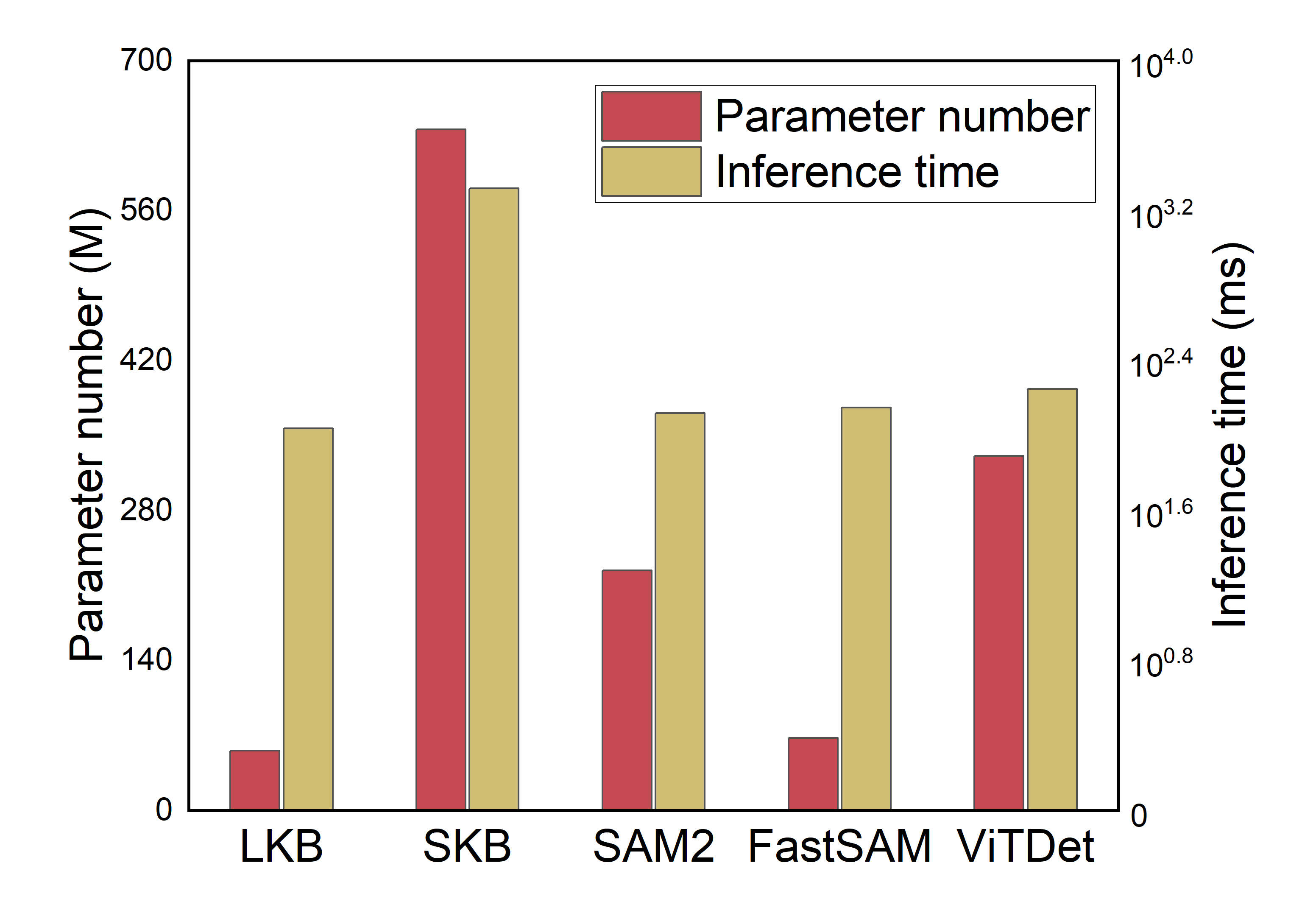}
	\caption{Comparison of LKB with other KBs and Vision Models.}
	\label{fig:fkbexp}
\end{figure}

The number of parameters is a crucial factor in determining whether a model can be deployed on devices with heterogeneous computational capabilities. Smaller parameter sizes lead to lower deployment costs for constructing the corresponding knowledge base, and they are particularly important for TokenCom, where the knowledge base must provide tokenizable spatial priors with minimal overhead. Moreover, TokenCom systems require low-latency perception-to-tokenization, and therefore shorter inference time for the knowledge base can significantly improve end-to-end TokenCom efficiency and responsiveness. We compare LKB with semantic knowledge base (SKB) \cite{jiang2024large} and representative alternatives, including SAM2 \cite{ravi2024sam2}, FastSAM \cite{zhao2023fast}, and ViTDet \cite{Li2022ExploringPV}.
The comparison results are shown in Fig. \ref{fig:fkbexp}.

As shown in Fig. \ref{fig:fkbexp}, LKB significantly outperforms the other models in terms of parameter size and inference time. This advantage is attributed to its use of a compact architecture while maintaining competitive localization/segmentation capability. Consequently, LKB yields the lowest construction and deployment cost and is best aligned with the stringent latency requirements of TokenCom, enabling fast prior generation to support downstream tokenization and transmission.

\subsection{\textcolor{black}{Performance Evaluation for ETC}}
The Peak Signal-to-Noise Ratio (PSNR) and the Structural Similarity Index Measure (SSIM) \cite{hore2010image} are widely used metrics for assessing image reconstruction quality. PSNR quantifies fidelity by measuring the MSE between the original and reconstructed images, where higher values indicate smaller distortion. SSIM evaluates perceptual similarity by jointly considering luminance, contrast, and structural consistency; values closer to 1 imply higher similarity \cite{hore2010image}.

\begin{figure}[htbp]
	\centering
	\subfigure[]{
		\includegraphics[width=8.5cm]{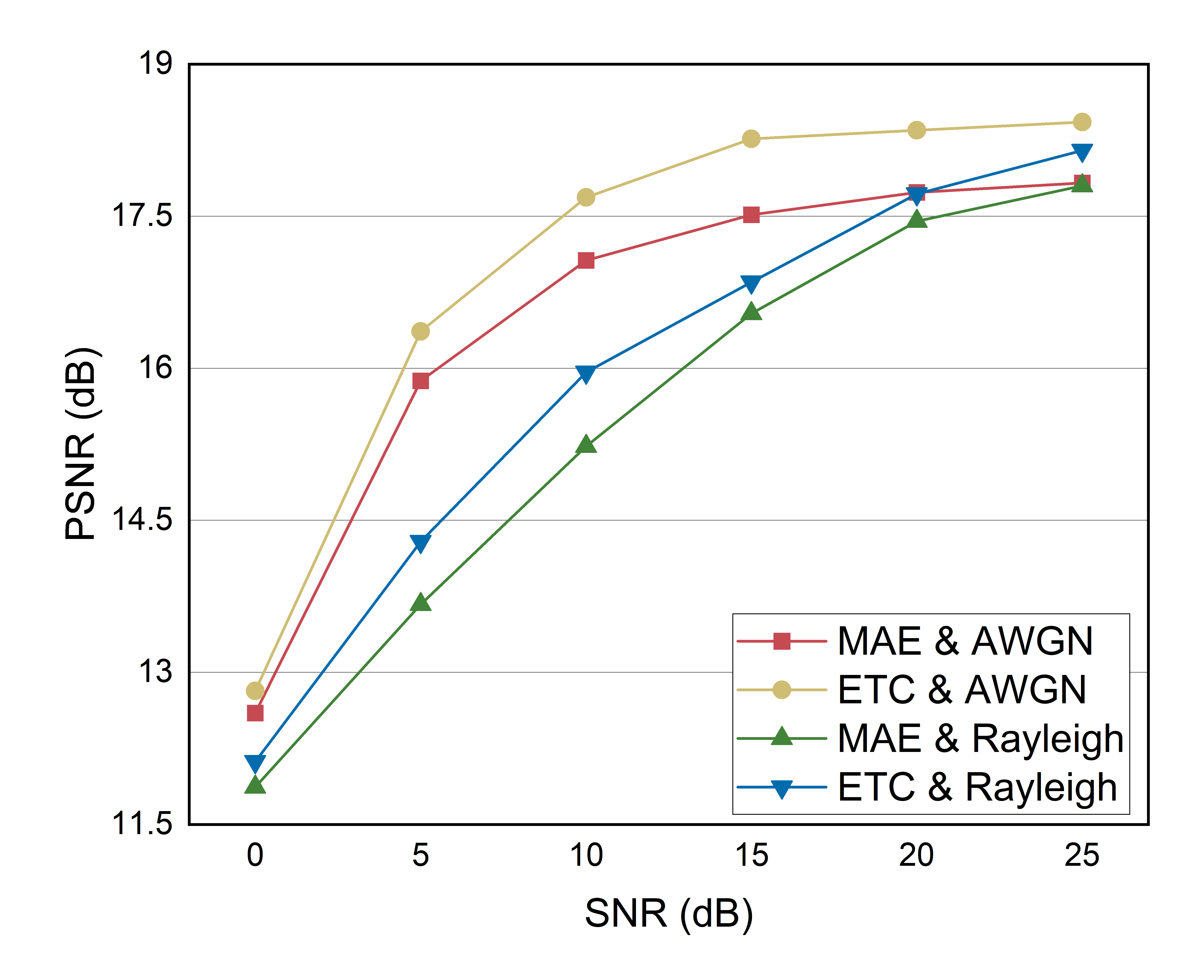}
		\label{fig:subfig1}
	}
	\quad
	\subfigure[]{
		\includegraphics[width=8.5cm]{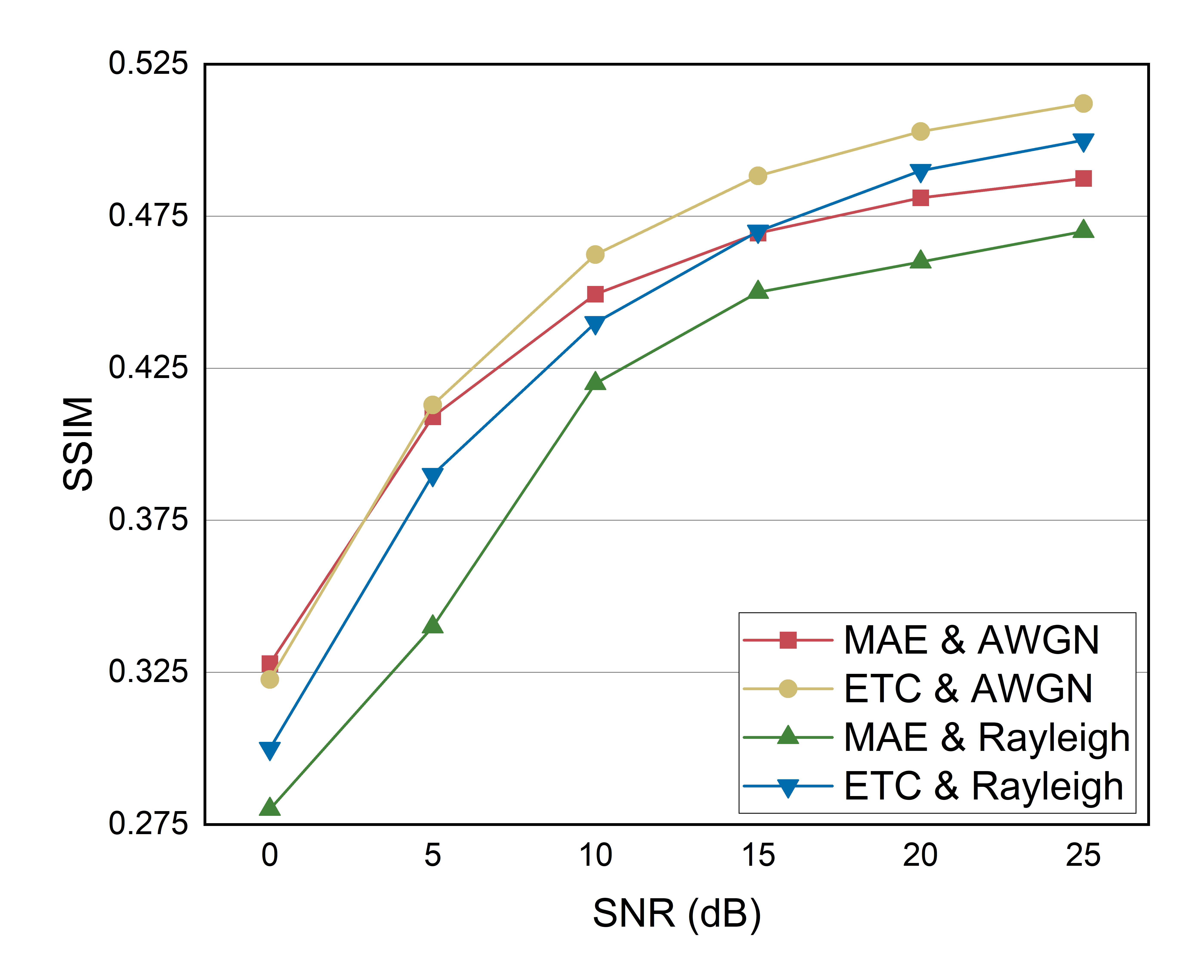}
		\label{fig:subfig2}
	}
	\caption{\textcolor{black}{Reconstruction Quality of ETC and MAE under AWGN and Rayleigh Fading.}}
	\label{fig:whole}
\end{figure}

We first compare PSNR and SSIM of the reconstructed content within the regions specified by the LKB under AWGN and Rayleigh fading channels. In this experiment, ETC and the standard MAE are respectively used as the encoder–decoder backbone. Since the LKB regions correspond to task-relevant objects, these metrics directly reflect reconstruction quality for the critical content that TokenCom aims to preserve. The results are reported in Fig. \ref{fig:whole}.

As shown in Fig. \ref{fig:subfig1} and Fig. \ref{fig:subfig2}, ETC achieves significantly higher PSNR and SSIM than MAE within the LKB-specified regions. This indicates that ETC not only inherits the efficiency of MAE-style masked encoding, but also improves the fidelity of object-centric reconstruction. The gain comes from two aspects: (i) ETC retains the compression advantage by encoding only a small subset of patches, and (ii) its LKB-guided adaptive masking allocates the limited encoding budget preferentially to object-related patches, thereby reducing the negative impact of background-dominated redundancy.

\begin{figure}[htbp]
	\centering
	\subfigure[]{
		\includegraphics[width=8.5cm]{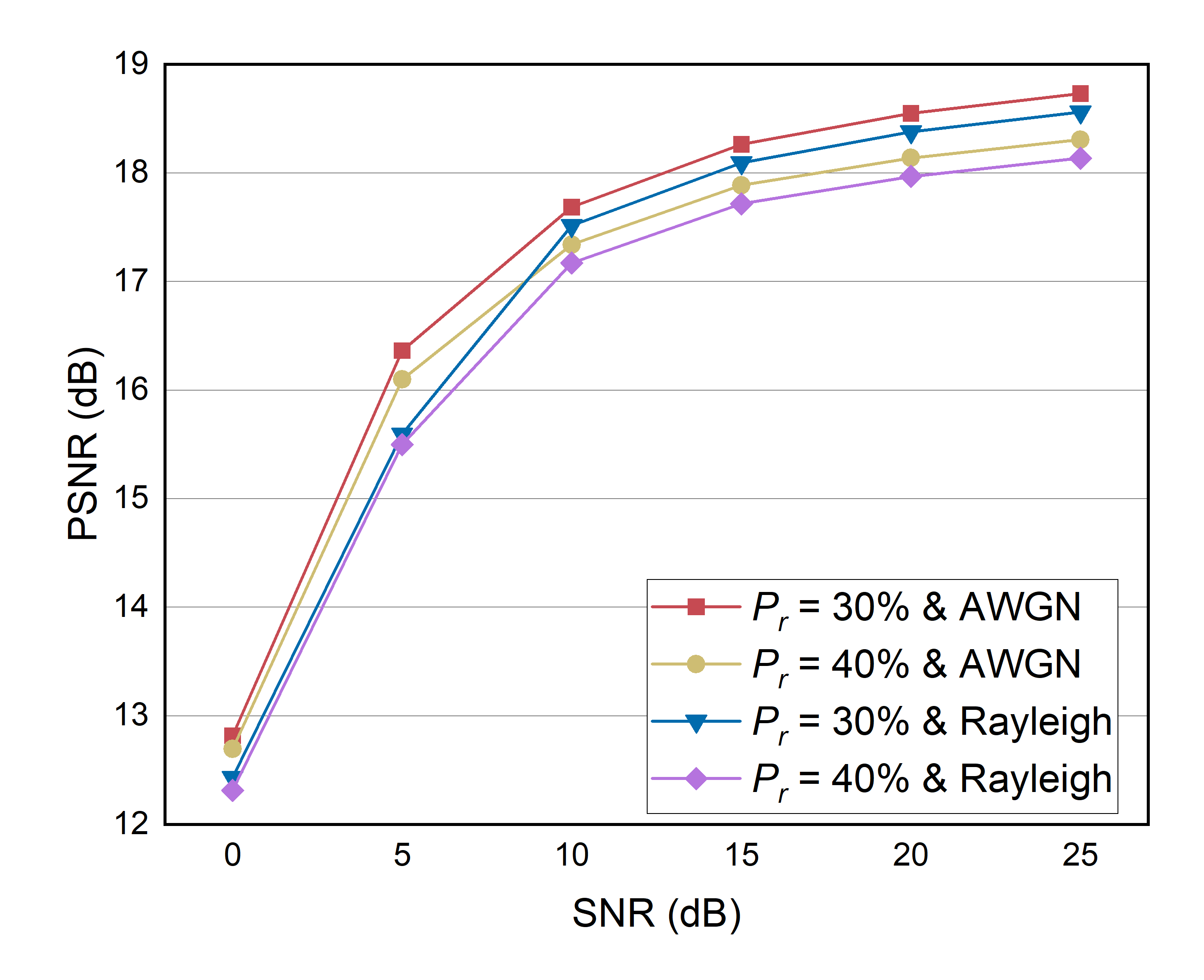}
		\label{fig:subfig3}
	}
	\quad
	\subfigure[]{
		\includegraphics[width=8.5cm]{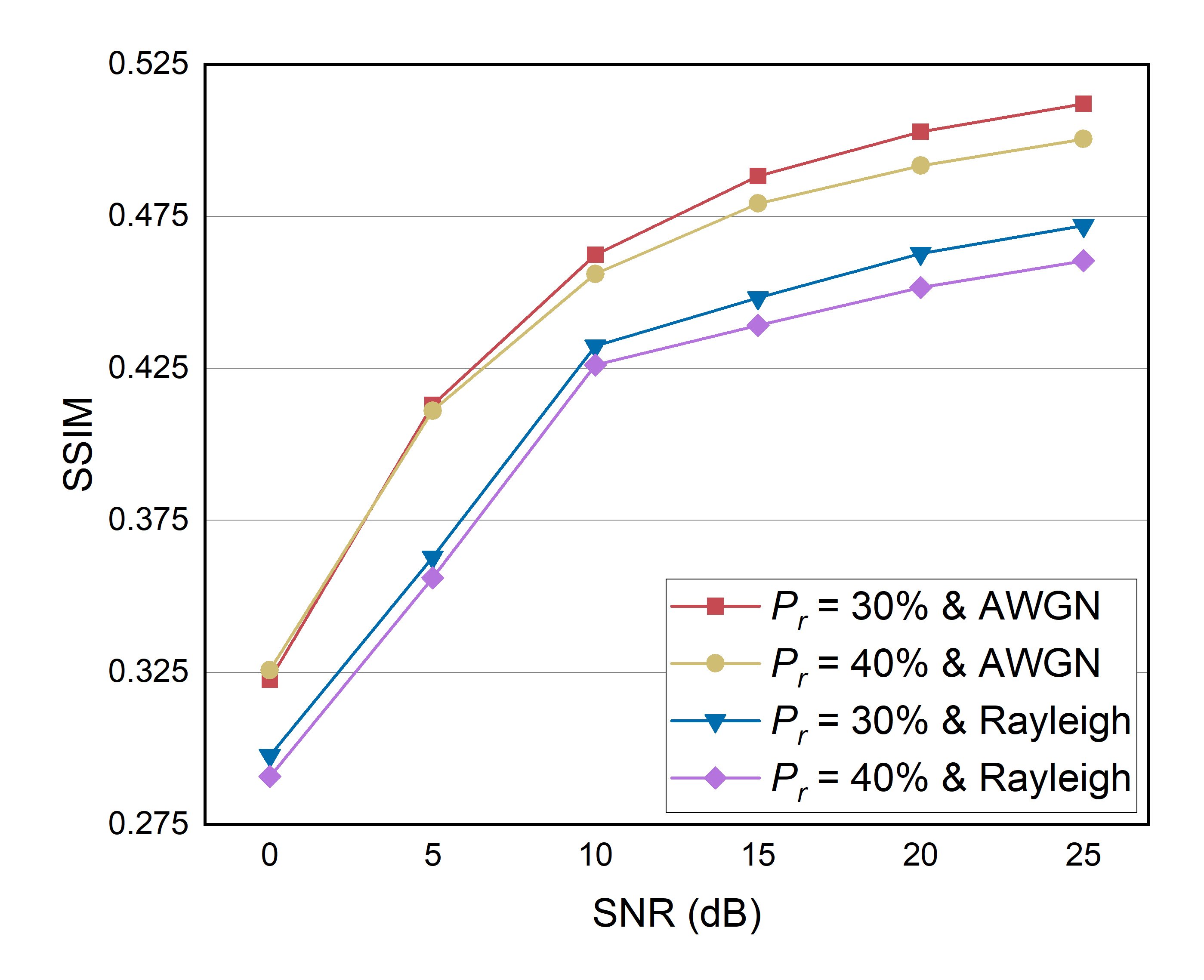}
		\label{fig:subfig4}
	}
	\caption{\textcolor{black}{Reconstruction Quality of ETC under Different Masking Probabilities.}}
	\label{fig:whole2}
\end{figure}

To further study the influence of the masking probability for object patches, denoted by \(P_r\), we conduct ablation experiments with different \(P_r\)
values under both AWGN and Rayleigh fading channels, as shown in Fig. \ref{fig:whole2}. As observed in Fig. \ref{fig:subfig3} and Fig. \ref{fig:subfig4}, the best reconstruction quality is achieved when 
\(P_r\) is set to 30\%. When 
\(P_r\) is increased beyond 30\% (e.g., 40\%), the reconstruction quality decreases, suggesting that an overly conservative masking strategy may dilute the encoding budget and weaken object-centric representation. Compared with the random masking in MAE, the proposed adaptive masking consistently improves reconstruction quality for object regions. Nevertheless, background patches cannot be completely ignored, and the residual dependence on background masking still affects the final reconstruction.

\textcolor{black}{Masking sensitivity also depends on object scale. Using a fixed $P_r$ is simple and effective, but it may deviate from the optimum when the target occupies an extreme fraction of the image. For very large targets, an excessively low $P_r$ (i.e., aggressive masking) may discard informative boundary/context patches and reduce reconstruction fidelity; for very small targets, an overly high $P_r$ tends to retain too many background patches, introducing redundancy and degrading transmission efficiency. This mild scale dependence is consistent with the empirical optimum around 30\% in Fig.~\ref{fig:subfig3} and Fig.~\ref{fig:subfig4}. In practice, a scale-aware schedule that maps the estimated target-area ratio to a narrow range (e.g., $P_r\in[0.2,0.4]$) could improve robustness.}

\subsection{Performance Evaluation for MTS Transmission}
\begin{figure}[htbp]
	\centering
	\includegraphics[width=8.5cm]{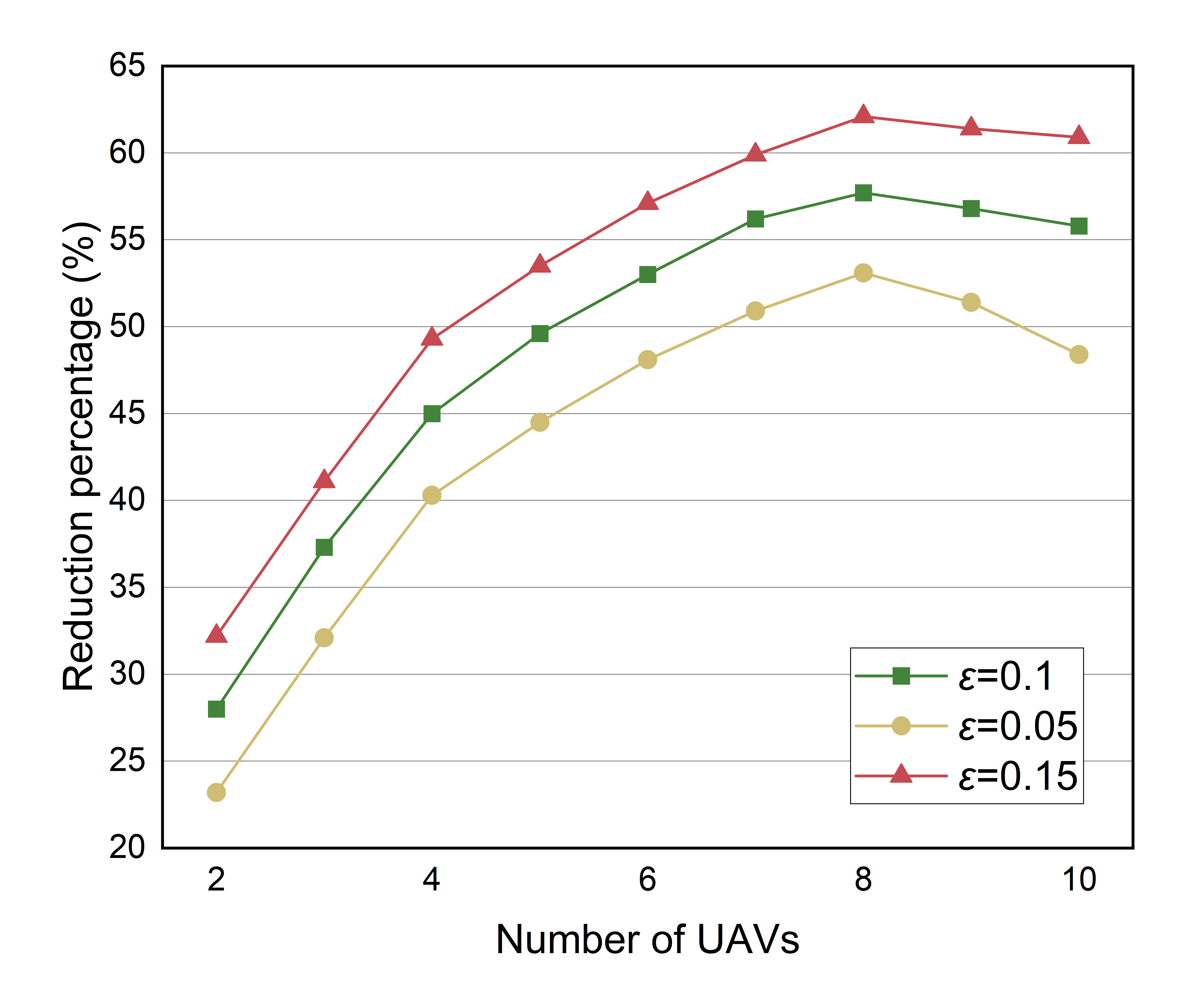}
	\caption{Impact of the Number of UAVs on Data Reduction Ratio.}
	\label{fig:mdmaexp}
\end{figure}

Public tokens commonly recur across different images, making bandwidth reduction in multi-UAV TokenCom systems a natural outcome of token reuse. However, as the number of UAVs increases, the definition of ``public'' tokens becomes stricter, because tokens that appear consistent among a small subset of UAVs may no longer remain consistent once new UAVs with divergent observations are included. As a result, the public-token set can shrink and the bandwidth savings may saturate or slightly decrease. In our simulations, we evaluate different thresholds $\epsilon$ for public-token identification and gradually increase the number of UAVs from $2$ to $10$. We report the data-reduction ratio, defined as the proportion of saved transmission data relative to the total transmitted data, and the results are shown in Fig.~\ref{fig:mdmaexp}.

As shown in Fig.~\ref{fig:mdmaexp}, under different $\epsilon$ settings, the data-reduction ratio generally increases as the number of UAVs grows, but the gain gradually saturates and may slightly drop when the swarm becomes large. Moreover, a larger $\epsilon$ relaxes the public-token criterion, leading to consistently higher reuse and a milder downward trend. In the early stage (small $K$), adding UAVs increases the chance that different images share common patterns, so more public tokens can be reused and the curve rises. As $K$ further increases, cross-UAV diversity becomes more pronounced, and a larger fraction of tokens are classified as private, which reduces the incremental benefit of sharing and can cause a slight decline in the reduction ratio.

\subsection{TokenCom Performance Evaluation}
\begin{figure}[htbp]
	\centering
	\subfigure[AWGN]{
		\includegraphics[width=8.5cm]{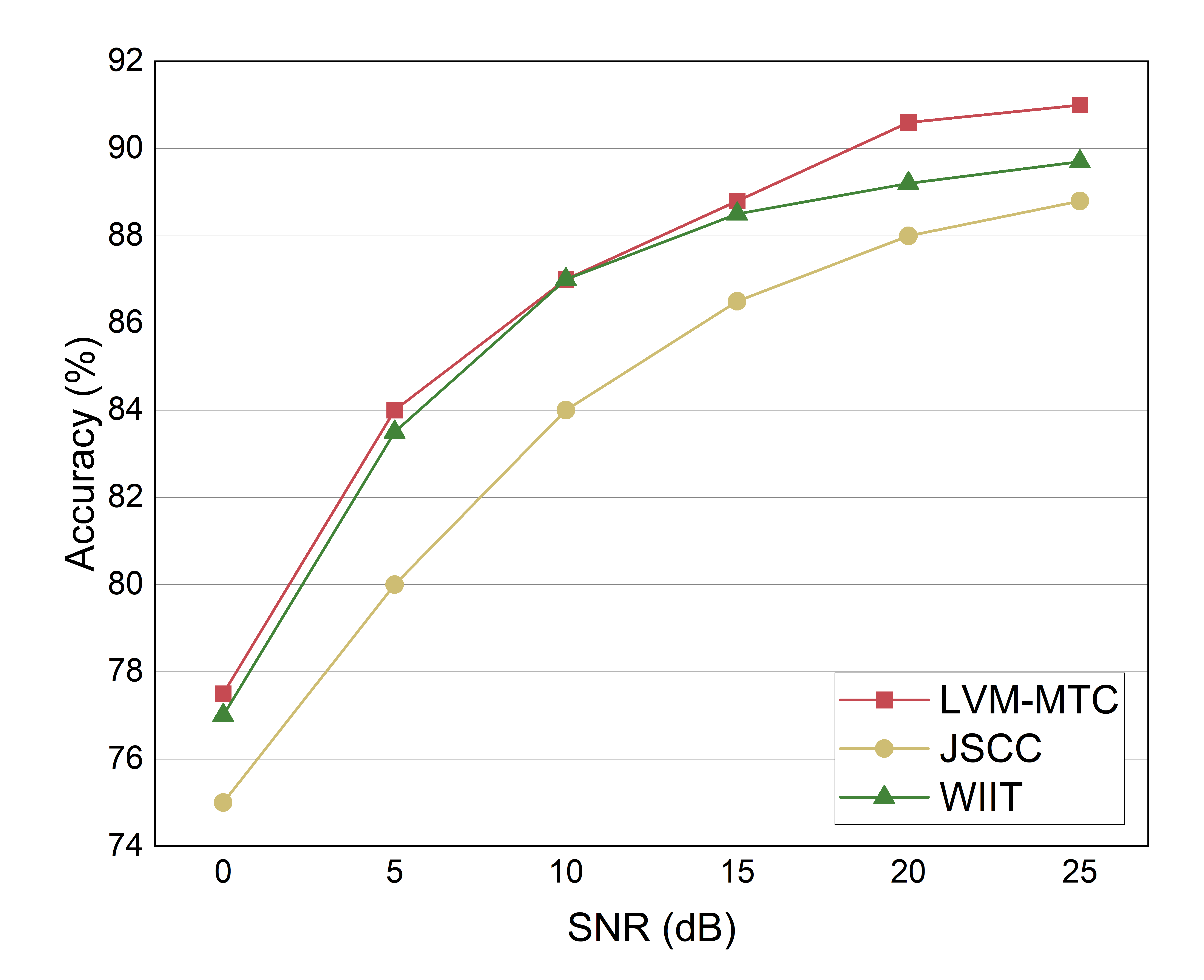}
		\label{fig:scawgn}
	}
	\quad
	\subfigure[Rayleigh Fading]{
		\includegraphics[width=8.5cm]{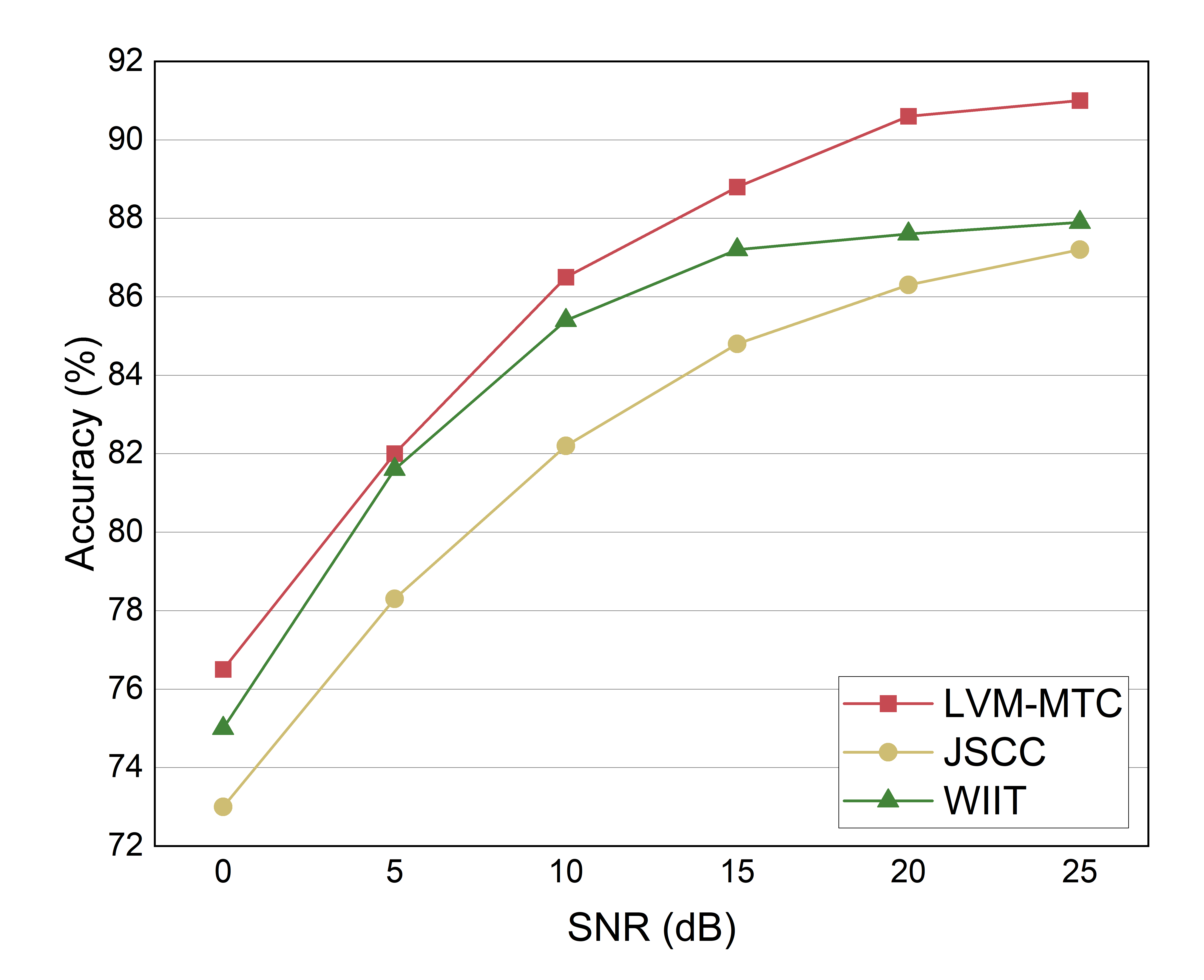}
		\label{fig:scfade}
	}
	\caption{\textcolor{black}{Performance Comparison of Different Methods for Classification.}}
	\label{fig:scexp}
\end{figure}

To evaluate the performance of the LVM-MTC system in image classification tasks, we compare it with two SC systems based on CNN (JSCC) \cite{bourtsoulatze2019deep} and ViT (WITT) \cite{yang2023witt} under both AWGN and Rayleigh fading channels. The image classification dataset used is CIFAR-10, and the performance metric for evaluation is classification accuracy. The experimental results are shown in Fig. \ref{fig:scexp}.

As shown in Fig. \ref{fig:scexp}, under both channel environments, LVM-MTC and WITT outperform JSCC in terms of classification accuracy. This is because both LVM-MTC and WITT utilize token/semantic encoders based on Transformer architectures, which enables them to extract image features more effectively than the CNN-based JSCC. Moreover, at higher SNR levels, LVM-MTC demonstrates superior performance over WITT. This can be attributed to the collaborative function of the token encoder and LKB in LVM-MTC, which allows the system to focus more on the main objects in the image while minimizing the impact of background noise and other irrelevant information.

\section{Conclusion}
This paper presents an innovative LVM-MTC system tailored for UAV-assisted low-altitude wireless networks. A key component of the system is the LKB, which is built upon a customized FlashSAM model capable of rapidly and accurately localizing key semantic objects in images available at the BS for UAV-specific transmission tasks. In addition, the system introduces an ETC based on the MAE architecture, which enables effective tokenization and compression by selectively encoding pixels in object-centric regions. Furthermore, we develop a multi-UAV token space and its corresponding MTS transmission strategy to support public/private token splitting and token reuse among multiple UAVs. Simulation results demonstrate the feasibility and effectiveness of the proposed LVM-MTC system.

From a system-deployment perspective, broadcasting public tokens may introduce privacy exposure risks by revealing sensitive object cues. Therefore, practical deployments should incorporate access control, token-level semantic filtering, and privacy-preserving tokenization mechanisms. Meanwhile, consistency of the lightweight LKB across multiple UAVs can be maintained through periodic model/metadata synchronization and drift-aware updates. Looking ahead, the proposed LKB–ETC pipeline and multi-UAV token space provide a transferable foundation for video and multimodal TokenCom, online knowledge/codec adaptation in dynamic environments, and agent-based UAV collaboration that coordinates cooperative perception, token routing, and resource allocation. Future work will further extend the system toward maritime and other cross-domain environments by investigating ocean-surface-related propagation effects and developing corresponding datasets.

\bibliographystyle{IEEEtran}
\bibliography{bare_jrnl_bobo}
\vspace{-1cm}
\begin{IEEEbiography}[{\includegraphics[width=1in,height=1.25in,clip,keepaspectratio]{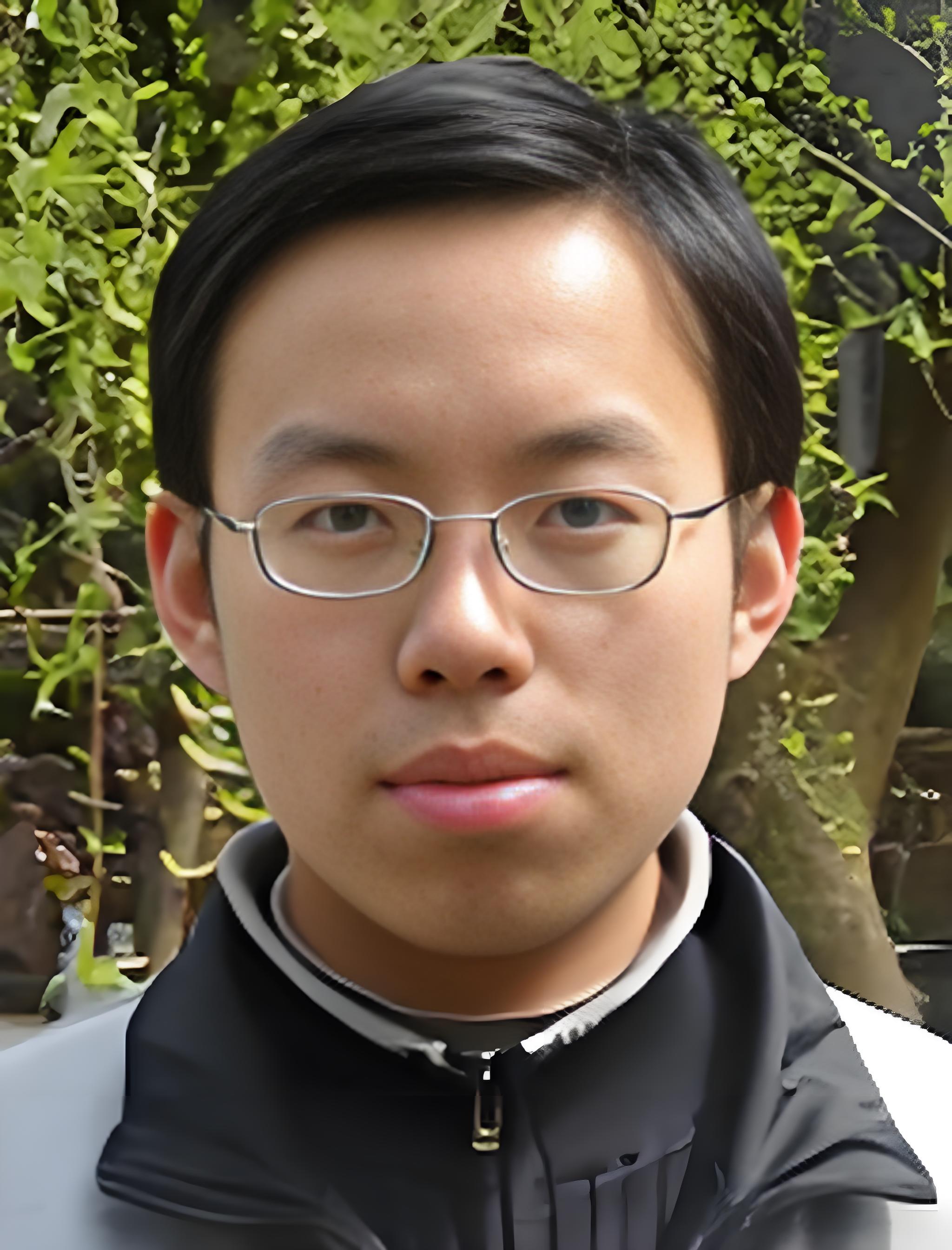}}]{Feibo Jiang} received his B.S. and M.S. degrees in School of Physics and Electronics from Hunan Normal University, China, in 2004 and 2007, respectively. He received his Ph.D. degree in School of Geosciences and Info-physics from the Central South University, China, in 2014. He is currently an associate professor at the Hunan Provincial Key Laboratory of Intelligent Computing and Language Information Processing, Hunan Normal University, China. His research interests include machine learning, Agentic AI, semantic communication, and mobile edge computing.
\end{IEEEbiography}
\vspace{-1cm}
\begin{IEEEbiography}[{\includegraphics[width=1in,height=1.25in,clip,keepaspectratio]{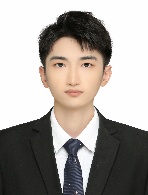}}]{Siwei Tu} received the B.S. degree from Jiangsu University of Technology, China, in 2023, where he is currently pursuing the master’s degree with the College of Information Science and Engineering at Hunan Normal University, China. His main research interests include federated learning and semantic communication.
\end{IEEEbiography}
\vspace{-1cm}
\begin{IEEEbiography}[{\includegraphics[width=1in,height=1.25in,clip,keepaspectratio]{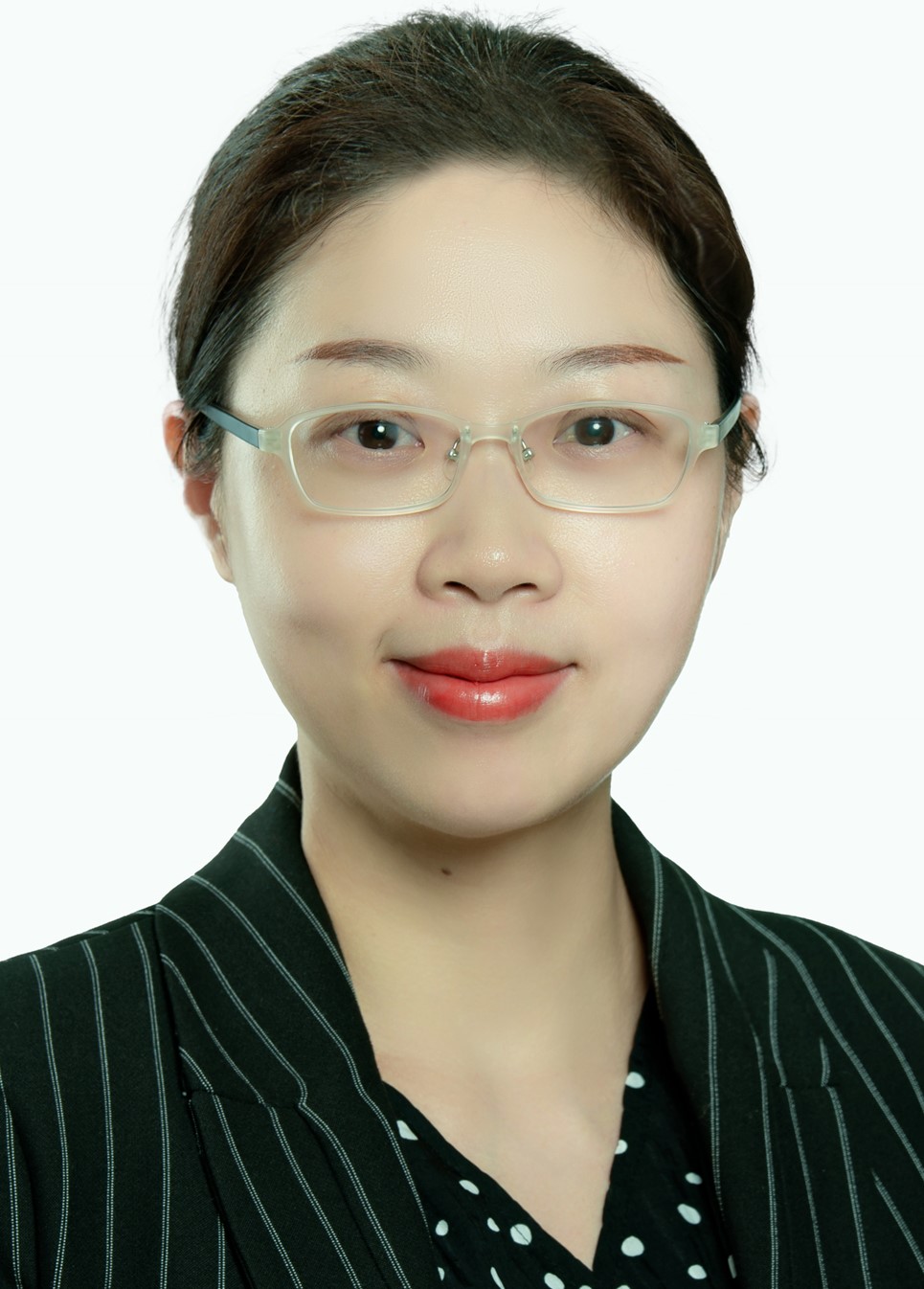}}]{Li Dong} received the B.S. and M.S. degrees in School of Physics and Electronics from Hunan Normal University, China, in 2004 and 2007, respectively. She received her Ph.D. degree in School of Geosciences and Info-physics from the Central South University, China, in 2018. She is currently a professor at Hunan University of Technology and Business, China. Her research interests include machine learning, Internet of Things, and mobile edge computing.
\end{IEEEbiography}
\vspace{-1cm}
\begin{IEEEbiography}[{\includegraphics[width=1in,height=1.25in,clip,keepaspectratio]{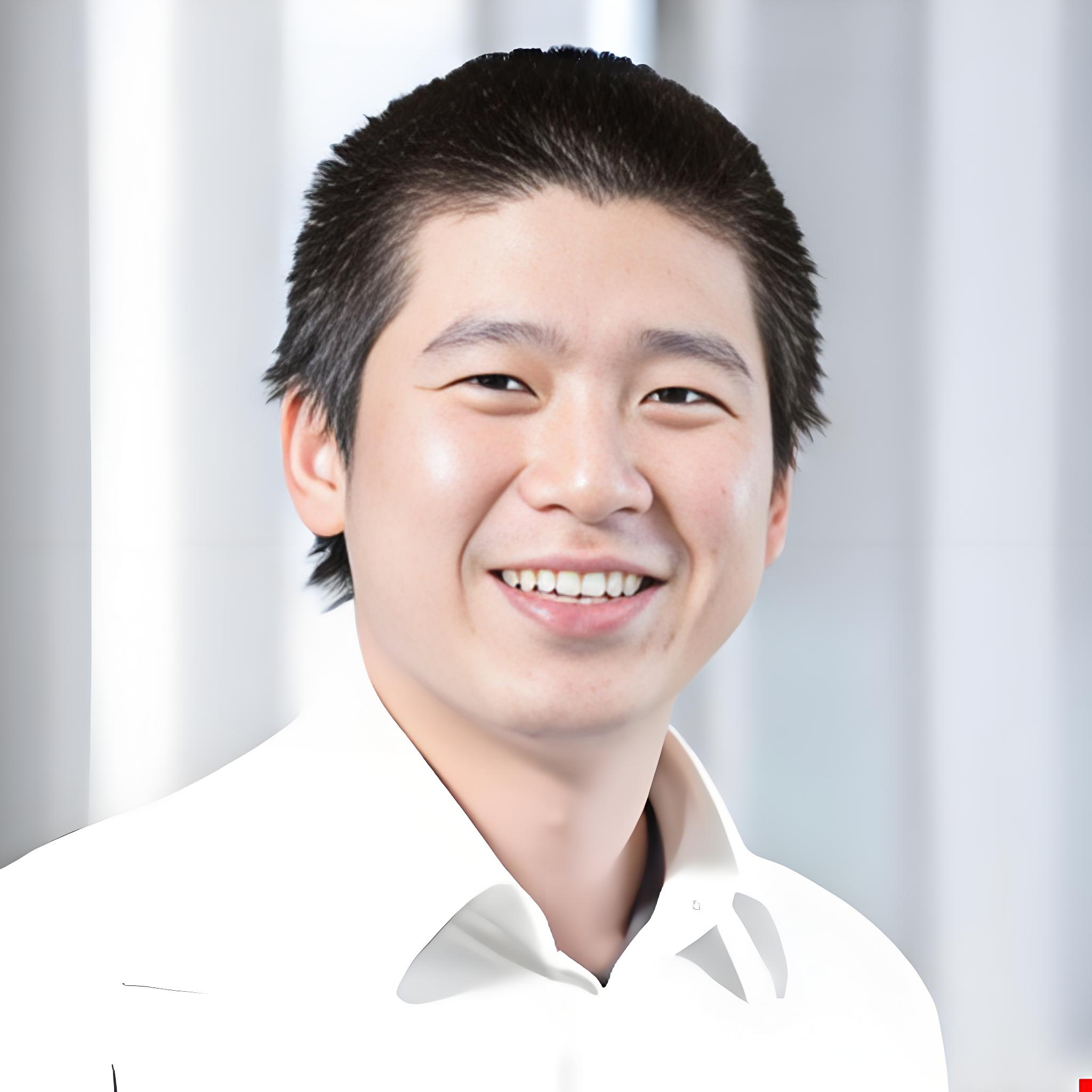}}]{Kezhi Wang} received the Ph.D. degree from the University of Warwick, U.K., funded by the Chancellor’s International Scholarship. He is currently a Professor with the Department of Computer Science, Brunel University London, U.K. He is also a Royal Society Industry Fellow and has been recognised as a Highly Cited Researcher by Clarivate Web of Science and a Top 2\% Scientist of the World. He has published over 150 papers in IEEE journals and has received several prestigious awards, including the IEEE Communications Society Leonard G. Abraham Prize, Heinrich Hertz Award, and Fred W. Ellersick Prize. His research interests include semantic communications, machine learning for communication systems, mobile edge computing, and wireless networks.
\end{IEEEbiography}
\vspace{-1cm}
\begin{IEEEbiography}[{\includegraphics[width=1in,height=1.25in,clip,keepaspectratio]{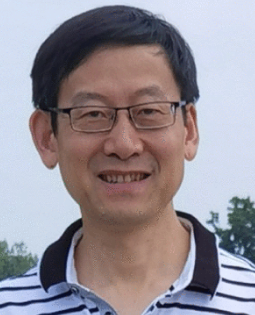}}]{Kun Yang} received his PhD from the Department of Electronic \& Electrical Engineering of University College London (UCL), UK. He is currently a Chair Professor of Nanjing University and also an affiliated professor at the University of Essex. His main research interests include wireless networks and communications, communication-computing cooperation, and new AI (artificial intelligence) for wireless. He has published 500+ papers and filed 50 patents. He serves on the editorial boards of a number of IEEE journals (e.g., IEEE WCM, TVT, TNB). He is a Deputy Editor-in-Chief of IET Smart Cities Journal. He has been a Judge of the GSMA GLOMO Award at World Mobile Congress – Barcelona since 2019. He was a Distinguished Lecturer of IEEE ComSoc (2020-2021), a Recipient of the 2024 IET Achievement Medals, and a Recipient of the 2024 IEEE CommSoft TC’s Technical Achievement Award. He is a Member of Academia Europaea (MAE), a Fellow of IEEE, a Fellow of IET, and a Distinguished Member of ACM.
\end{IEEEbiography}
\vspace{-1cm}
\begin{IEEEbiography}[{\includegraphics[width=1in,height=1.25in,clip,keepaspectratio]{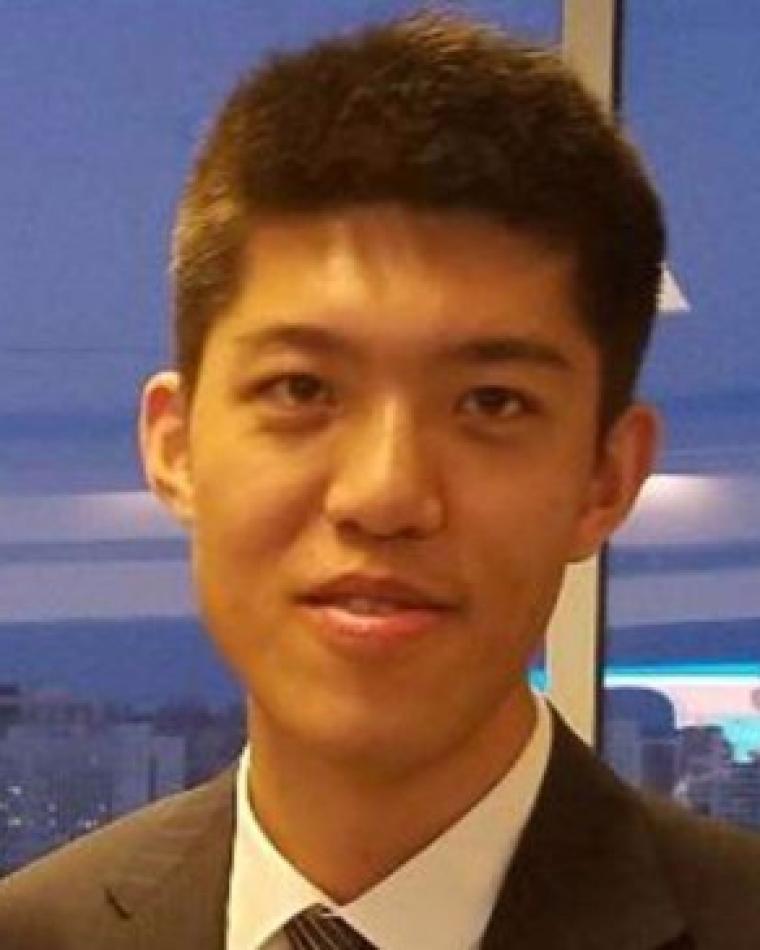}}]{Ruiqi Liu} is a master researcher in the Wireless and Computing Research Institute of ZTE Corporation, responsible for long-term research as well as standardization. His main research interests include reconfigurable intelligent surfaces, integrated sensing and communication and 6G requirements. He is the author or co-author of several books and book chapters. He has made significant contributions to standardization of 5G / 5G-advanced in 3GPP by authoring and submitting more than 500 technical documents with over 100 approved, and serving as a rapporteur. He served as the chair of multiple correspondence and drafting groups in ITU-R WP5D towards 6G. He currently serves as the Vice Chair of ISG RIS in the ETSI. He is involved in organizing committees of international conferences and is invited to give multiple talks, including the keynote speech at IEEE Globecom 2024. He is the Voting Member of the ComSoc Industry Communities Board and the ComSoc Standards Development Board. He served as the Deputy Editor-in-Chief of IET Quantum Communication, the Associate Editor for IEEE Communications Letters, the Associate Editor for IEEE Communications Magazine and Guest Editor for a series of special issues. His recent awards include the 2023 Beijing Science and Technology Invention Award, 2025 IEEE SPCC Early Achievement Award and the Best Paper Award from the Intelligent and Converged Networks (2025). He is listed in the World's Top 2\% Scientists by Stanford/Elsevier.
\end{IEEEbiography}
\vspace{-1cm}
\begin{IEEEbiography}[{\includegraphics[width=1in,height=1.25in,clip,keepaspectratio]{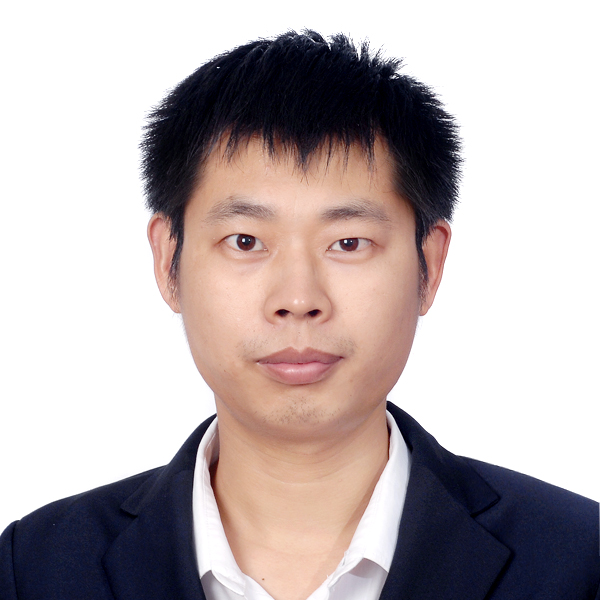}}]{Cunhua Pan} received the B.S. and Ph.D. degrees from the School of Information Science and Engineering, Southeast University, Nanjing, China, in 2010 and 2015, respectively. He was a Research Associate with the University of Kent, Canterbury, U.K., from 2015 to 2016. He held a postdoctoral position with the Queen Mary University of London, London, U.K., from 2016 and 2019, and was a Lecturer from 2019 to 2021. He has been a Full Professor with Southeast University since 2021. He has published over 120 IEEE journal papers. His research interests mainly include reconfigurable intelligent surfaces (RIS), intelligent reflection surface, ultrareliable low-latency communication, machine learning, UAV, Internet of Things, and mobile-edge computing.
\end{IEEEbiography}
\vspace{-1cm}
\begin{IEEEbiography}[{\includegraphics[width=1in,height=1.25in,clip,keepaspectratio]{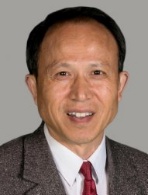}}]{Jiangzhou Wang} is currently a Professor with Southeast University, Nanjing, China. He has published more than 500 papers and five books, with research interests primarily in mobile communications. He is an International Member of the Chinese Academy of Engineering (CAE) and a Fellow of the Royal Academy of Engineering (RAEng), U.K. He received the 2024 IEEE Communications Society Fred W. Ellersick Prize and the 2022 IEEE Communications Society Leonard G. Abraham Prize. He also served as the Technical Program Chair of the 2019 IEEE International Conference on Communications (ICC 2019) in Shanghai, the Executive Chair of IEEE ICC 2015 in London, and the Technical Program Chair of IEEE WCNC 2013.
\end{IEEEbiography}
\newpage

\end{document}